\documentclass[aps,prd,twocolumn,superscriptaddress,preprintnumbers,nofootinbib]{revtex4-2}%
\usepackage{epsf,epsfig}
\usepackage[utf8]{inputenc}
\usepackage{amssymb,amsmath,amsfonts}
\usepackage{color}
\usepackage{slashed}
\usepackage{tensor} 
\usepackage{braket}
\usepackage{array}
\usepackage{float}
\usepackage[colorlinks=true]{hyperref}  
\hypersetup{
    bookmarks=true,         % show bookmarks bar?
    unicode=false,          % non-Latin characters 
    pdftoolbar=true,        % show Acrobat
    pdfmenubar=true,        % show Acrobat 
    pdffitwindow=false,     % window fit to page when opened
    pdfstartview={FitH},    % fits the width of the page to the window
    colorlinks=true,       % false: boxed links; true: colored links
    linkcolor=magenta, %red,          % color of internal links (change box color with linkbordercolor)
    citecolor=blue,        % color of links to bibliography
    filecolor=magenta,      % color of file links
    urlcolor=cyan           % color of external links
} 
\newcommand{\bea}{\begin{eqnarray}}

\newcommand{\eea}{\end{eqnarray}}
\usepackage{lipsum} 
\newcommand{\ba}{\begin{eqnarray}}
\newcommand{\ea}{\end{eqnarray}}

\newcommand{\beq}{\begin{equation}}
\newcommand{\eeq}{\end{equation}}
\newcommand{\beqa}{\begin{eqnarray}}
\newcommand{\eeqa}{\end{eqnarray}}
\newcommand{\beqar}{\begin{eqnarray*}}
\newcommand{\eeqar}{\end{eqnarray*}}

\newcommand{\E}{\mathcal{E}}

\usepackage{color}

\newcommand{\dvtag}{\\ &}
\newcommand{\dvvtag}{\right.\\ &\left.}

\newcommand{\req}[1]{(\ref{#1})} %{Eq.\thinspace(\ref{#1})}

\begin{document}

\allowdisplaybreaks

%\title{Black hole quasinormal modes in higher-derivative gravity: rotation, overtones and harmonics}

\title{Higher-derivative corrections to the Kerr quasinormal mode spectrum}

\author{Pablo A. Cano}
\email{pablo.cano@icc.ub.edu}
\affiliation{Departament de F\'isica Qu\`antica i Astrof\'isica, Institut de Ci\`encies del Cosmos\\
 Universitat de Barcelona, Mart\'i i Franqu\`es 1, E-08028 Barcelona, Spain}
 
 \author{Lodovico Capuano}
\email{lcapuano@sissa.it}
\affiliation{SISSA, Via Bonomea 265, 34136 Trieste, Italy and INFN Sezione di Trieste}
\affiliation{IFPU - Institute for Fundamental Physics of the Universe, Via Beirut 2, 34014 Trieste, Italy}

 \author{Nicola Franchini}
\email{franchini@apc.in2p3.fr}
\affiliation{Universit\'e Paris Cit\'e, CNRS, Astroparticule et Cosmologie, F-75013 Paris, France}
\affiliation{CNRS-UCB International Research Laboratory, Centre Pierre Bin\'etruy,
IRL2007, CPB-IN2P3, Berkeley, CA 94720, USA}

 \author{Simon Maenaut}
\email{simon.maenaut@kuleuven.be}
\affiliation{Institute for Theoretical Physics, KU Leuven. Celestijnenlaan 200D, B-3001 Leuven, Belgium}

 \author{Sebastian H. V\"olkel}
\email{sebastian.voelkel@aei.mpg.de}
\affiliation{Max Planck Institute for Gravitational Physics (Albert Einstein Institute),
D-14476 Potsdam, Germany}

\date{\today}

\begin{abstract}
We provide the most complete analysis so far of quasinormal modes of rotating black holes in a general higher-derivative extension of Einstein's theory.
By finding the corrections to the Teukolsky equation and expressing them in a simple form, we are able to apply a generalized continued fraction method that allows us to find the quasinormal mode frequencies including overtones. We obtain the leading-order corrections to the Kerr quasinormal mode frequencies of all the $(l,m,n)$ modes with $l=2,3,4$, $-l\le m\le l$ and $n=0,1,2$, and express them as a function of the black hole spin $\chi$ using polynomial fits. We estimate that our results remain accurate up to spins between $\chi\sim 0.7$ and $\chi\sim 0.95$, depending on the mode.  We report that overtones are overall more sensitive to corrections, which is expected from recent literature on this topic. We also discuss the limit of validity of the linear corrections to the quasinormal mode frequencies by estimating the size of nonlinear effects in the higher-derivative couplings. All our results are publicly available in an online repository. 
\end{abstract} 

\maketitle

\section{Introduction}
By detecting the gravitational waves (GWs) produced by compact binary mergers, we can learn about fundamental physics \cite{Berti:2018cxi,Berti:2018vdi,Barack:2018yly,Barausse:2020rsu}. In particular, black hole binaries allow us to test vacuum general relativity (GR) in a highly warped and dynamical regime in which the theory has never been probed before. This advance in the experimental state of the art compels us to not only test Einstein's theory, but also to look for signatures of new physics, as GW observations may unveil new effects which would otherwise remain hidden. 

One of the purest tests of GR that would reveal the presence of new physics is black hole spectroscopy \cite{Kokkotas:1999bd,Dreyer:2003bv,Berti:2009kk,Konoplya:2011qq,Berti:2018vdi,Barack:2018yly,Silva:2022srr,Franchini:2023eda,Maselli:2023khq,Liu:2024atc}, consisting in the identification of the quasinormal modes (QNMs) of the post-merger black hole during the ringdown phase.  QNMs are damped sinusoid modes with characteristic frequencies and damping times that control the response of a black hole after it is perturbed. In GR, black holes are described by the Kerr metric, and as a consequence of the no-hair theorem, the full spectrum of QNMs is univocally determined by the black hole mass and angular momentum. Thus, the measurement of more than one mode provides a test of GR. On the other hand, modifications of GR lead to deviations with respect to the Kerr QNM spectrum that could be spotted with ringdown observations. 

Motivated by the possibility of placing bounds on extensions of GR, there has been a growing interest in the literature in computing the QNMs of black holes in theories of gravity beyond GR, \textit{e.g.} \cite{Cardoso:2009pk,Blazquez-Salcedo:2016enn,Cardoso:2018ptl,deRham:2020ejn,Cano:2020cao,Moura:2021eln,Pierini:2021jxd,Wagle:2021tam,Srivastava:2021imr,Bryant:2021xdh,Cano:2021myl,Pierini:2022eim,Miguel:2023rzp,Silva:2024ffz,Chung:2024ira,Chung:2024vaf,Blazquez-Salcedo:2024oek}. However, this is a very challenging problem, especially for rotating black holes.  Rotation is an important feature, as the post-merger black hole formed in binary collisions typically has large angular momentum, with a probability distribution that peaks around a value of the rotation parameter $\chi\equiv a/M \sim 0.7$ \cite{Fishbach:2017dwv}. 
In GR, the study of perturbations of Kerr black holes is made possible by the Teukolsky equation \cite{Teukolsky:1973ha}, a second order, decoupled and separable equation that arises thanks to several special features of the Einstein's equations and the Kerr metric (\textit{i.e.}, Petrov type D and existence of a Killing tensor \cite{Walker:1970un}). The loss of these properties in extensions of GR makes the study of perturbations of modified Kerr black holes an extremely challenging problem. 

A significant progress has been recently achieved thanks to the development of generalized Teukolsky equations  \cite{Li:2022pcy,Hussain:2022ins,Cano:2023tmv,Wagle:2023fwl} that hold for arbitrary extensions of Einstein gravity. This has led to the first computation ever of QNMs of modified Kerr black holes with substantial angular momentum in Ref.~\cite{Cano:2023jbk}. Other approaches based on spectral methods are also showing promise in obtaining the QNMs of black holes with large angular momentum beyond GR \cite{Blazquez-Salcedo:2023hwg,Chung:2024ira,Chung:2024vaf,Blazquez-Salcedo:2024oek}. 

In this work, we take on the modified Teukolsky approach to provide the most complete computation yet of QNMs of rotating black holes beyond GR. We consider a general effective field theory (EFT) extension of Einstein's theory with up to eight-derivative terms \cite{Endlich:2017tqa,Cano:2019ore,Ruhdorfer:2019qmk}, which we treat as perturbative corrections to GR. Thus, our analysis is limited to small deviations but it has the advantage of being very general: the EFT captures the most general modification of GR as long as there are no additional massless degrees of freedom besides the graviton. 
Furthermore, the assumption of small deviations is well motivated based on the fact that the current GW observations are consistent with GR. 

Our analysis extends that of \cite{Cano:2023jbk} in several ways, the most important one being the computation of overtones. By combining the modified Teukolsky equation predicted by the higher-derivative theories together with the continued fraction method of \cite{Cano:2024jkd}, we obtain for the first time the corrections to the overtones of rotating black holes. In addition, \cite{Cano:2023jbk} only studied the $(l,m,n)=(2,2,0)$ and $(3,3,0)$ modes, and here we extend this to all the modes with $l=2,3,4$, $-l\le m\le l$ and $n=0,1,2$.  We also provide a more detailed analysis on the structure of the modified Teukolsky equations and on the convergence of the spin expansion that we utilize to find them. Thanks to including more terms in this expansion and making a sensible use of Pad\'e approximants, we manage to obtain results for angular momentum as high as $\chi=0.95$ for some modes. Finally, we also provide an estimation of the validity of the first-order correction to the QNM frequencies.

The paper is organized as follows. In Section~\ref{sec:EFT} we review the EFT extension of GR and its rotating black hole solutions, expressed as a first-order correction in the higher-derivative coupling constants and as a series expansion in the angular momentum. In Section~\ref{sec:Teuk} we review the procedure of \cite{Cano:2023tmv} to obtain the modified Teukolsky radial equations and we reduce them to a simple form in terms of a potential that depends on four coefficients. We then study some of their properties and the convergence of the spin expansion. We obtain the corrections to the Kerr QNMs from these modified Teukolsky equations in Section~\ref{sec:QNMs}, which we express as polynomials in the black hole angular momentum. We discuss several features of our results, like the convergence of the spin expansion, the growth of the overtones and the size of non-linear effects in the coupling. We use cubic gravity to illustrate these discussions but our full results are available in the public repository \cite{gitbeyondkerr}. 
Finally, we conclude in Section~\ref{sec:conclusions}, where we discuss future directions. Throughout the paper we work in units with $G=c=1$. 

% \cite{Hirano:2024fgp}

\section{The EFT extension of General Relativity}\label{sec:EFT}
Following the logic of EFT, the Einstein-Hilbert action is considered to be simply the leading order term in an effective action for the gravitational interaction. In general, this effective action will contain an infinite tower of higher-derivative terms, which --- if we assume general covariance --- must be built from contractions of the Riemann tensor and its covariant derivatives.   At each order in the higher-derivative expansion, one should include all possible terms of that type in order to capture a general effective action. Thus, one gets an action of the form
\begin{equation}
S_{\rm EFT}=\frac{1}{16\pi}\int d^4x\sqrt{|g|}\bigg[R+\sum_{n=2}^{\infty} \ell^{2(n-1)}\mathcal{L}_{(n)} \bigg]\, ,
\end{equation}
where $\mathcal{L}_{(n)}$ denotes, schematically, a general linear combination of all the Lagrangian densities with $2n$ derivatives of the metric, and $\ell$ is the length scale that determines the cutoff of the effective theory.\footnote{For simplicity in this presentation we are assuming that all the corrections appear at the same scale, but it could also happen that different terms appear at different scales, \textit{e.g.} because they arise due to different mechanisms. } 
However, one is also able to perform redefinitions of the metric tensor of the form
\begin{equation}
g_{\mu\nu}\rightarrow g_{\mu\nu}+\ell^2 \Delta^{(2)}_{ \mu\nu}+\ell^4 \Delta^{(4)}_{ \mu\nu}+\ldots\, ,
\end{equation}
where $\Delta^{(2n)}_{ \mu\nu}$ are symmetric rank-2 tensors with $2n$ derivatives. This redefinition changes some terms in the effective action, allowing one to cancel many of them. In particular, one can show that any term containing Ricci curvature can be removed in this form, and hence the action be reduced to a sum of Weyl invariants. 

Field redefinitions deserve a word of caution. Although the physics in the original and the transformed frame are \textit{equivalent} (in the sense that there is a map between them), they are not necessarily the \textit{same}. That is, the transformation of the metric could affect some of the physical (and observable) properties of the spacetime. However, it turns out that many properties of black holes actually remain invariant under metric redefinitions. Among them, black hole thermodynamic properties \cite{Jacobson:1993vj}, multipole moments \cite{Cano:2022wwo} and quasinormal mode frequencies \cite{deRham:2020ejn} are invariant. Thus, for the purpose of this paper we can use field redefinitions to simplify the effective action and capture the effects of a very general theory in a minimal set of parameters. 

One finds that, to eight derivatives\footnote{For a discussion of the EFT corrections at arbitrary order, see \cite{Ruhdorfer:2019qmk}.}, the most general EFT extension of GR can always be written in the form \cite{Endlich:2017tqa,Cano:2019ore}
\begin{equation}\label{eq:EFT}
\begin{aligned}
S_{\rm EFT}=\frac{1}{16\pi}\int& d^4x\sqrt{|g|}\bigg[R+\ell^4\left(\lambda_{\rm ev}\mathcal{R}^3+\lambda_{\rm odd}\tilde{\mathcal{R}}^3\right)\\
+&\ell^6\left(\epsilon_{1}\mathcal{C}^2+\epsilon_{2}\tilde{\mathcal{C}}^2+\epsilon_{3}\mathcal{C}\tilde{\mathcal{C}}\right)+\mathcal{O}(\ell^8) \bigg]\, ,
\end{aligned}
\end{equation}
with higher-curvature invariants
\begin{align*}
\mathcal{R}^3&=\tensor{R}{_{\mu\nu }^{\rho\sigma}}\tensor{R}{_{\rho\sigma }^{\delta\gamma }}\tensor{R}{_{\delta\gamma }^{\mu\nu }}\, ,& \tilde{\mathcal{R}}^3&=\tensor{R}{_{\mu\nu }^{\rho\sigma}}\tensor{R}{_{\rho\sigma }^{\delta\gamma }} \tensor{\tilde R}{_{\delta\gamma }^{\mu\nu }}\, ,\\
\mathcal{C}&=R_{\mu\nu\rho\sigma} R^{\mu\nu\rho\sigma}\, ,&
\tilde{\mathcal{C}}&=R_{\mu\nu\rho\sigma} \tilde{R}^{\mu\nu\rho\sigma}\, , %\\
%&\phantom{\rm{where}} \, &\tilde{R}_{\mu\nu\rho\sigma}&=\frac{1}{2}\epsilon_{\mu\nu\alpha\beta}\tensor{R}{^{\alpha\beta}_{\rho\sigma}} \, .
\end{align*}
and where
\begin{equation}
 \tilde{R}_{\mu\nu\rho\sigma}=\frac{1}{2}\epsilon_{\mu\nu\alpha\beta}\tensor{R}{^{\alpha\beta}_{\rho\sigma}}
 \end{equation}
is the dual Riemann tensor. Due to the presence of $\tilde{R}$, the terms with $\lambda_{\rm odd}$ and $\epsilon_{3}$ violate parity and have interesting implications for the QNM spectrum. 
The coefficients $\lambda_{\rm ev, odd}$, $\epsilon_{1,2,3}$ are dimensionless coupling constants, since the scale of the corrections is set by the common length scale $\ell$. Otherwise one can absorb $\ell$ into these coefficients and work with dimensionful coupling constants. We remark that this action contains no four-derivative terms since they are trivial: they can always be expressed as a combination of $R^2$, $R_{\mu\nu}R^{\mu\nu}$ --- which do not modify the vacuum GR solutions and can be removed by field redefinitions --- and the Gauss-Bonnet density $\mathcal{X}_{4}=R_{\mu\nu\rho\sigma}R^{\mu\nu\rho\sigma}-4R_{\mu\nu}R^{\mu\nu}+R^2$ --- which is topological and does not affect the equations of motion.

The modified Einstein's equations take the form
\begin{equation}\label{EinsteinEqTeff}
G_{\mu \nu}+\ell^4\E^{(6)}_{\mu\nu}+\ell^6\E^{(8)}_{\mu\nu}=0\, ,
\end{equation}
where
\begin{equation}\label{Tmunun}
\E^{(n)}_{\mu\nu}=\tensor{P}{^{(n)}_{\mu}^{\rho \sigma \gamma}} \tensor{R}{_{\nu \rho \sigma \gamma}}-\frac{1}{2}g_{\mu \nu} \mathcal{L}_{(n)}+2 \nabla^\sigma \nabla^\rho P^{(n)}_{\mu \sigma\nu\rho}\, ,
\end{equation}
and the tensor $P^{(n)}_{\mu\nu\rho \sigma}$ is the partial derivative of the corresponding Lagrangian with respect to the Riemann tensor, which yields\footnote{We note that, when evaluated on a Ricci flat spacetime, these expressions can be simplified by using that $\tilde{R}_{\mu\nu\rho\sigma}=\tilde{R}_{\rho\sigma\mu\nu}$.}
\begin{align}
\label{eq:Pcubic}
&P^{(6)}_{\mu\nu\rho \sigma}=3\lambda_{\rm ev} \tensor{R}{_{\mu\nu}^{\alpha \beta}}\tensor{R}{_{\alpha\beta\rho\sigma}}\\\notag
&+\lambda_{\rm odd}\left(\tensor{R}{_{\mu\nu}^{\alpha \beta}}\tensor{\tilde R}{_{\alpha\beta\rho\sigma}}+\tensor{R}{_{\mu\nu}^{\alpha \beta}}\tensor{\tilde R}{_{\rho\sigma\alpha\beta}}+\tensor{R}{_{\rho\sigma}^{\alpha \beta}}\tensor{\tilde R}{_{\mu\nu\alpha\beta}}\right)\, ,\\
\label{eq:Pquartic}
&P^{(8)}_{\mu\nu\rho \sigma}=4\epsilon_1 \mathcal{C} R_{\mu\nu\rho \sigma}+2 \epsilon_2\mathcal{\tilde C}\left(\tilde R_{\mu\nu\rho \sigma}+\tilde R_{\rho \sigma\mu\nu}\right)\\\notag
&+\epsilon_{3}\left[2\mathcal{\tilde C}R_{\mu\nu\rho \sigma}+\mathcal{C}\left(\tilde R_{\mu\nu\rho \sigma}+\tilde R_{\rho \sigma\mu\nu}\right)\right]\, .
\end{align}

The magnitude of higher-curvature corrections depends on the curvature scale of the spacetime. In the case of a black hole, the curvature scale around the horizon is set by the black hole mass $M$, and it is thus useful to introduce the dimensionless couplings
\begin{equation}\label{alphaq}
\alpha_{\rm ev}=\frac{\ell^4\lambda_{\rm ev}}{M^4}\, ,\quad \alpha_{\rm odd}=\frac{\ell^4\lambda_{\rm odd}}{M^4}\, ,\quad \alpha_{i}=\frac{\ell^6\epsilon_{i}}{M^6}\, ,
\end{equation}
which characterize the magnitude of the relative corrections to GR. In order to be within the regime of validity of the EFT, we require that these couplings be small $|\alpha_{\rm q} |\ll 1$, where ${\rm q} \in \left\lbrace {\rm ev, \rm odd, 1, 2, 3}\right\rbrace$. 
Assuming this, we will work perturbatively at first order in the $\alpha_{\rm q}$ couplings, which should be an accurate approximation as long as the couplings are small enough.  However, exactly how small these couplings need to be to ensure the validity of the leading order perturbative expansion depends a great of deal on which quantity one is computing. In the case of QNMs, it has recently been shown by \cite{Silva:2024ffz} that the breakdown of the linear regime happens much earlier for overtones than for fundamental modes --- in fact, for high enough overtone index, the EFT always breaks down. One expects a similar phenomenon for eikonal modes. Here we explore these questions in the presence of rotation.

\subsection{Rotating black holes}  The higher-derivative terms in the effective action \eqref{eq:EFT} affect the rotating black hole solutions of theory, which are no longer given by the Kerr metric. In the perturbative regime, the rotating black hole solutions can be conveniently written in the form \cite{Cano:2019ore}
\begin{align}\label{eq:ansatz}
    ds^2 =& -\left(1-\frac{2M r}{\Sigma}-H_1\right)dt^2\\\notag
    &
    -(1+ H_2)\frac{4a M r (1-x^2)}{\Sigma}dtd\phi
    \\\notag
    &+\left(1+H_3\right)\Sigma\left(\frac{dr^2}{\Delta}+\frac{dx^2}{1-x^2}\right)\\\notag
    &+(1+H_4)\left(r^2 + a^2+\frac{2a^2Mr(1-x^2)}{\Sigma}\right)(1-x^2) d\phi^2\, , %\\
\end{align}
where 
\begin{equation}
\Delta=r^2-2Mr+a^2\, ,\quad \Sigma=r^2+a^2x^2\, ,
\end{equation}
and $x=\cos\theta$.  In this way, the four functions $H_i(r,x)$ parametrize the corrections to the Kerr metric expressed in Boyer-Lindquist coordinates, which we recover when $H_i(r,x)=0$. One interesting aspect about this ansatz is that the location of the horizon remains unchanged in terms of the radial coordinate $r$, since it corresponds to the largest root of $\Delta$:
\begin{equation}
r_{+}=M+\sqrt{M^2-a^2}\, .
\end{equation}
Furthermore, we impose suitable boundary conditions on the $H_i$ functions such that the parameters $M$ and $a$ have the same meaning as in the Kerr solution: the total mass and the angular momentum per unit mass, respectively.  

At leading order in the corrections, the $H_i$ functions receive a linear contribution from each of the higher-derivative terms in \eqref{eq:EFT} 
\begin{equation}
H_i=\sum_{\rm q} \alpha_{\rm q}H_{i,\rm q}+\mathcal{O}(\alpha_{\rm q}^2)\, .
\end{equation}
However, determining the $H_{i,\rm q}$ functions by solving the corrected Einstein's equations is a complicated problem, and so far no exact solutions have been found. One possibility consists in solving these equations numerically, as in \textit{e.g.} \cite{Kleihaus:2011tg,Delsate:2018ome,Horowitz:2024dch}, although this approach has not been applied so far for the theories in \eqref{eq:EFT}. 
Here we follow \cite{Cano:2019ore} and find the solution as a series expansion in the dimensionless spin parameter
\begin{equation}
\chi=a/M\, .
\end{equation}
As shown by \cite{Cano:2019ore}, at each order in the spin expansion one can find the solution for the $H_i$ functions analytically, and in general the result takes the form
\begin{equation}
\label{eq:H-exp}
    H_{i,\rm q} = \sum_{n=0}^{\infty}\chi^n  \sum_{p=0}^{n} \sum_{k=0}^{k_\text{max}(n)}H_{i, \rm q}^{(n,p,k)}\left(\frac{M}{r}\right)^k x^p \, ,
\end{equation}
where each term is a polynomial in $x$ and $1/r$ with coefficients $H_{i, \rm q}^{(n,p,k)}$ that are determined analytically. This series is convergent in the black hole exterior for $\chi<1$, which means that one can in principle study any sub-extremal black hole in this form. 
However, the convergence is quite slow and many terms are needed to obtain an accurate solution for highly spinning black holes. In this paper we use expansions of order $\chi^{18}$, which remain accurate up to spins $\chi\sim 0.8$, although the convergence can be faster depending on the observable \cite{Cano:2023qqm}. 

It is worth mentioning that, despite the exact form of the $H_i$ functions still being unknown, some properties of the corrected Kerr black holes can actually be obtained analytically. For instance some of the multipole moments \cite{Cano:2022wwo} as well as thermodynamic quantities \cite{Reall:2019sah} such as entropy, angular velocity and temperature (surface gravity) are known analytically. One can use the Taylor expansion of those exact expressions as a check of the convergence of the spin expansion.

\section{The modified Teukolsky equations}\label{sec:Teuk}
The study of perturbations of rotating black holes of arbitrary angular momentum in extensions of GR has only been made possible thanks to the generalizations of Teukolsky equations of Refs.~\cite{Li:2022pcy,Hussain:2022ins,Cano:2023tmv}. Here we use the ``universal Teukolsky equations'' of \cite{,Cano:2023tmv}, although all the proposals should be equivalent. We refer to that work for details, but let us explain here the basic idea behind these modified Teukolsky equations.  

These are equations for the gravitational perturbations over the background \req{eq:ansatz}, expressed using Newman-Penrose (NP) variables. 
Schematically, at first order in the higher-derivative couplings, these modified Teukolsky equations take the form 
\begin{align}\label{eq:modTeukPsi}
\mathcal{O}^{(0)}_{+2}(\delta \Psi_{0})+\alpha  \mathcal{O}^{(1)}_{+2}(\delta \Psi_{n},\delta e^{a}, \delta \gamma_{abc})&=0\, ,\\
\mathcal{O}^{(0)}_{-2}(\delta \Psi_{4})+\alpha  \mathcal{O}^{(1)}_{-2}(\delta \Psi_{n},\delta e^{a}, \delta \gamma_{abc})&=0\, ,
\end{align}
where $\alpha$ denotes one of the couplings in \req{alphaq} and where $\delta \Psi_{n}$, $\delta e^{a}$ and $\delta \gamma_{abc}$  are respectively, the perturbation of the Weyl scalars, NP frame and spin connection. The objects $\mathcal{O}^{(n)}_{\pm 2}$ are linear differential operators, where $n=0$ corresponds to the Teukolsky operator. Thus, when $\alpha=0$ these equations reduce to the usual Teukolsky equations for the variables $\delta \Psi_{0}$ and $\delta \Psi_{4}$. For $\alpha\neq 0$ the equations are no longer decoupled, but one can achieve a decoupled equation by expressing $\delta \Psi_{n}$, $\delta e^{a}$ and $\delta \gamma_{abc}$  in terms of the Teukolsky variables $\delta \Psi_{0}$ and $\delta \Psi_{4}$ by means of a metric reconstruction. Since we only work at first order in $\alpha$, it suffices to use the result of metric reconstruction of Kerr perturbations in GR \cite{Wald:1978vm}. We review this process in the appendix \ref{app:perturbations}, since it contains some details that are relevant to understand the structure of the resulting equations. 
An important aspect of it is that the full problem requires us to consider not only the variables $\delta \Psi_{0}$ and $\delta \Psi_{4}$, but also their NP conjugates $\delta \Psi_{0}^{*}$ and $\delta \Psi_{4}^{*}$, that satisfy modified Teukolsky equations similar (but different) to \req{eq:modTeukPsi}. The reason for including the NP conjugate variables is that we work with a complex metric perturbation, and therefore the NP conjugates no longer represent complex conjugation, but new independent variables.

Metric reconstruction allows us to get decoupled equations which are nevertheless nonseparable.  In order to deal with this, we decompose the Teukolsky variables in spin-weighted spheroidal harmonics $S^{lm}_{s}(x;a\omega)$ as
\begin{equation}\label{deltapsiexpansion}
\begin{aligned}
\delta \Psi_0&=e^{-i\omega t+i m \phi} \sum_{l}  R^{lm}_{+2}(r)S^{lm}_{+2}(x;a\omega)\, ,\\
\delta \Psi_4&=e^{-i\omega t+i m \phi}  \sum_{l} \zeta^{-4}R^{lm}_{-2}(r)S^{lm}_{-2}(x;a\omega)\, ,\\
\delta \Psi_0^{*}&=e^{-i\omega t+i m \phi} \sum_{l}R^{*lm}_{2}(r)S^{lm}_{-2}(x;a\omega)\, ,\\
\delta \Psi_4^{*}&=e^{-i\omega t+i m \phi}  \sum_{l} (\zeta^{*})^{-4}R^{*lm}_{-2}(r)S^{lm}_{+2}(x;a\omega)\, ,
\end{aligned}
\end{equation}
where $\zeta=r-i a x$, 
and project the equations onto $S^{{l'}m}_{s}(x;a\omega)$, reducing them to an infinite system of coupled radial equations. We remark that here all the spin-weighted spheroidal harmonics are evaluated at the same frequency $\omega$, and thus one can use the known orthonormality property
\begin{equation}
2\pi \int dx S_{s}^{lm}(x, a\omega)S_{s}^{l'm}(x, a\omega)=\delta_{ll'}\, .
\end{equation}
This is different than when one considers the set of spheroidal harmonics with each one evaluated on its own QNM frequency $\omega_{lmn}$, since that set is not orthogonal  \cite{London:2020uva}. 

Then, a key observation is to note that the QNMs will be composed of a dominant $l$-mode in the sums \req{deltapsiexpansion}, say $l=l_0$, while the rest of the terms will be of order $\alpha$. This is because when $\alpha=0$, QNMs consist of a single term. 
Thus, using the orthogonality of the spin-weighted spheroidal harmonics, one obtains that the equations for the dominant-mode radial variables $R^{l_0m}_{\pm2}$, $R^{*l_0m}_{\pm2}$ are decoupled from the rest of the $l$-modes at first order in $\alpha$ \cite{Cano:2020cao,Ghosh:2023etd,Cano:2023tmv}.   By performing a series expansion in the spin $\chi$, these equations can be obtained in a fully analytic form. 
We study them next.

\subsection{Radial Teukolsky equations}
As a result of the computation we just sketched, we get four equations for the radial variables that read \cite{Cano:2023tmv}

\begin{equation}\label{radialeqs}
\begin{aligned}
\mathfrak{D}_{s}^2R_{s}-\alpha \left[ f_{s} R_{s}+ g_{s} \Delta \frac{dR_{s}}{dr}\right]=&0\, ,\\
\mathfrak{D}_{s}^2R^{*}_{s}-\alpha \left[ f^{*}_{s} R^{*}_{s}+ g^{*}_{s} \Delta \frac{dR^{*}_{s}}{dr}\right]=&0\, ,
\end{aligned}
\end{equation}
where $s=\pm 2$, we are removing the $lm$ labels for clarity,  and
\begin{equation}\label{Dsopdef}
\mathfrak{D}_{s}^2=\Delta^{-s+1}\frac{d}{dr}\left[\Delta^{s+1}\frac{d}{dr}\right]+V_{s} 
\end{equation}
is the Teukolsky operator with a potential $V_s$ given by 
\begin{equation}
\begin{aligned}
V_s&=(am)^2+\omega^2 \left(a^2+r^2\right)^2-4 a m M r \omega\\
&+i s \left(2 a m (r-M)-2 M \omega \left(r^2-a^2\right)\right)\\
&+\Delta\left(-a^2 \omega^2+s-B_{lm}+2 i r s \omega\right)\, ,
\end{aligned}
\end{equation}  
and $B_{lm}$ are the angular separation constants for the usual spin-weighted spheroidal harmonics, in the conventions of \cite{Cano:2023tmv}, such that $B_{lm}(a\omega=0)=l(l+1)-s^2$.  
In addition, $f_{s}$ and $g_{s}$ are functions of $r$ which, in the context of the spin expansion we are performing, take the form
\begin{equation}\label{fsgs}
\begin{aligned}
f_{s}(r)&=r^4\sum_{n=0}^{n_{\rm max}}\chi^n\sum_{k=0}^{k_{\rm max }(n)} \frac{f_{s,n,k} }{r^{k}}\, ,\\
g_{s}(r)&=r^3\sum_{n=0}^{n_{\rm max}}\chi^n\sum_{k=0}^{k_{\rm max }(n)} \frac{g_{s,n,k} }{r^{k}}\, ,
\end{aligned}
\end{equation}
for certain coefficients $f_{s,n,k}$, $g_{s,n,k}$ that we determine analytically. Analogous expressions hold for $f_{s}^{*}$,  $g_{s}^{*}$, and we remark that in general $f_{s}^{*}\neq f_{s}$, $g_{s}^{*}\neq g_{s}$ and that the star $*$ does not represent complex conjugation. 

These coefficients depend on the frequency $\omega$ and on a number of parameters that relate the different radial functions when $\alpha=0$ and that we use to fully decouple the corrected equations. In the Kerr case, the radial variables with $s=+2$ and $s=-2$ are related by the Starobinsky-Teukolsky (ST) identities \cite{teukolsky1972rotating,teukolsky1972rotating,Starobinsky:1973aij,Chandrasekhar:1984siy,Fiziev:2009ud}, that depend on two constants $C_{\pm 2}$ whose product is fixed --- see the appendix \ref{app:perturbations}. Thus, one of these constants is arbitrary, and it turns out to correspond to a gauge choice for the metric reconstruction. Physics should therefore be independent of this choice and we will come back to this later. This is a manifestation of the fact that the  $s=+2$ and $s=-2$ variables contain the same information \cite{Wald:1973wwa}. 
On the other hand, when $\alpha=0$, the $R_{s}$ variable and its NP conjugate $R_{s}^{*}$ satisfy the same equation and same boundary conditions, so that it follows that these variables must be proportional, 
\begin{equation}
R_{s}^{*}=q_{s}R_{s}\, .
\end{equation}
Contrary to the Starobinsky-Teukolsky constants, the $q_{s}$ parameters are physical as they determine the polarization of the perturbation. The reason why they do not show up in the study of Kerr perturbations has to do with isospectrality. But in theories beyond GR isospectrality is typically broken, and hence the corrections to the Teukolsky equations depend on the polarization --- see \cite{Li:2023ulk} for a study of isospectrality breaking in the Newman-Penrose formalism.
We find that the correction of the Teukolsky equation, captured by the functions \req{fsgs}, depends explicitly on the ST constants $C_{s}$ and the polarization parameters $q_{s}$. The dependence on these parameters has a particular form that we address below, after we make a few simplifications in the corrected Teukolsky equations.   

In order to simplify \req{radialeqs} we can perform a change of variables of the form
\begin{equation}\label{eq:changeofvariables}
\begin{aligned}
R_{s}&\rightarrow R_{s}+\alpha \left(A_{s}R_{s}+B_{s}\Delta \frac{dR_{s}}{dr}\right)\, ,\\
\end{aligned}
\end{equation}
where the functions $A_{s}$ and $B_{s}$ are linear in the higher-derivative couplings.  At first order in $\alpha$, this has the following effect on the corrected Teukolsky equation \req{radialeqs}: $f_{s}\rightarrow \tilde f_s$, $g_{s}\rightarrow \tilde{g}_{s}$, where 
\begin{align}\label{eq:fs}
\tilde f_{s}&=f_{s}-\Delta \left(\Delta  A_s''+(s+1) A_s' \Delta '-2 V_s B_s'-B_s V_s'\right)\, ,\\\label{eq:gs}
\tilde g_{s}&=g_{s}-\Delta\left(2 A_s'+\Delta B_s''-(s-1) B_s' \Delta '-2 s B_s\right)\, .
\end{align}
Then, we can choose the functions $A_s$ and $B_s$ so that the corrected equation takes a simple form. In particular, we can fix $\tilde g_{s}=0$, and determine $A_s$ from this condition.  With this choice, we rewrite the equation \req{radialeqs} as 
\begin{equation}\label{eq:correctedradial} 
\begin{aligned}
\Delta^{-s+1}\frac{d}{dr}\left[\Delta^{s+1}\frac{dR_{s}}{dr}\right]+\left(V_s+ \alpha \delta V_s\right) R_{s}=&0\, ,
\end{aligned}
\end{equation}  
where the correction to the potential, $\delta V_s$, reads
\begin{equation}\label{eq:deltaVsdef}
\begin{aligned}
\delta V_{s}=&-f_s+\frac{1}{2} \Delta g_s'+ s g_s \Delta'+B_s \Delta \left(-V_s'+s (1+s)
   \Delta' \right)\\
   &+\frac{1}{2} \Delta B_s' \left(-4 V_s+2(2s-1) \Delta+\left(s^2-1\right)
   \Delta'^2\right)\\
   &-\frac{3}{2} \Delta^2 \Delta' B_s''-\frac{1}{2} \Delta^3 B^{(3)}_{s}\, .
\end{aligned}
\end{equation}
We still have a functional freedom in the form of the potential $\delta V_{s}$, corresponding to the choice of $B_{s}(r)$. The quasinormal mode frequencies and other physical properties are nevertheless independent of this choice, as long as the function $B_s(r)$ is well behaved (\textit{e.g.}, it must decay fast enough at infinity and be regular at the horizon).
For convenience, let us from now on focus on the case of $s=-2$, although the discussion for the $s=2$ is analogous. 

In the context of the spin expansion and taking into account the expressions \req{fsgs}, it is also natural to assume a function $B_{-2}$ of the form
\begin{equation}
B_{-2}=\sum_{n=0}^{n_{\rm max}}\sum_{k=-1}^{k_{\rm max}}\frac{\chi^n}{r^k} b_{n,k}\, .
\end{equation}
Observe that we start the sum over $k$ at $k=-1$, corresponding to a linear term in $r$. This is the highest power of $r$ that preserves the asymptotic structure of the potential (which goes as $r^4$). On the other hand, this function is clearly regular at the horizon. With this choice, the potential \req{eq:deltaVsdef} also takes the form of a simultaneous expansion in $\chi$ and in $1/r$.
\begin{equation}
\delta V_{-2}=\sum_{n=0}^{n_{\rm max}}\sum_{k=-4}^{\tilde{k}_{\rm max}}\frac{\chi^n}{r^k} v_{n,k}\
\end{equation}

However, by tuning the coefficients $b_{n,k}$, we find that it is possible to remove almost all of the terms in this potential. All of the powers $1/r^k$ with $k>2$ can be removed. That is, we can set $v_{n,k}=0$ $\forall k\ge 3, n\ge 0$. Interestingly, the term with $1/r^2$ cannot be removed. The remaining freedom in the $b_{n,k}$ coefficients can be used to impose additional constraints on the terms with $k=-1,0,1,2,3,4$. We can for instance choose $v_{n,1}=0$, so that there is no $1/r$ term, and we can also set $v_{n,-4}=0$, so that the corrections to the potential are subleading at infinity. Interestingly, this choice also leads to $v_{n,-3}=0$. Once we have set these conditions, we find that there is no additional freedom and the potential is uniquely reduced to the simple form
\begin{equation}\label{eq:dVm2}
\delta V_{-2}=\frac{A_{-2}}{r^{2}}+A_0+A_1 r+A_2 r^2\, ,
\end{equation}
Thus, it is determined by the four coefficients $A_k$, which are expressed as a power expansion in the spin
\begin{equation}\label{eq:Akcoefficients}
A_k=\sum_{n=0}^{n_{\rm max}}\chi^n A_{k,n}\, .
\end{equation}
We have checked that for all the theories considered and all the different $l,m$ modes one can always write the potential as in \req{eq:dVm2}. In addition, since the potential can always be brought to this form at any order in the spin expansion, we expect that \req{eq:dVm2} may be general result that applies even non-perturbatively in the spin. 

We have analytically computed the coefficients $A_{k}$ at order $n_{\rm max}=18$ for the $l=2,3$, $-l\le m\le l$ modes and at order $n_{\rm max}=14$ for $l=4$, $-l\le m\le l$, for all the theories in \req{eq:EFT}. These expressions are too lengthy to be displayed in the text, but we provide them in the \texttt{github} repository \cite{gitbeyondkerr}.

\subsection{Properties of the potentials}
As we mentioned earlier, corrections to the Teukolsky equation depend on the Starobinsky-Teukolsky constants $C_{s}$, representing a gauge choice in the metric reconstruction, and on the polarization parameters $q_{s}$. The dependence of the correction to the potential $\delta V_{s}$ (and of the functions $f_{s}$ , $g_{s}$ we started with) on these parameters is of the form
  \begin{equation}\label{eq:deltaVCq}
 \begin{aligned}
 \delta V_{s}&=\frac{1}{P_{s}}\left[\delta V_{s}^{(1)}+q_{s}\delta V_{s}^{(2)}+C_{s}\delta V_{s}^{(3)}+C_{s}q_{-s}\delta V_{s}^{(4)}\right]\, ,\\
 \delta V_{s}^{*}&=\frac{1}{q_{s}P_{s}^{*}}\left[\delta V_{s}^{*(1)}+q_{s}\delta V_{s}^{*(2)}+C_s \delta V_{s}^{*(3)}+q_{-s}C_{s}\delta V_{s}^{*(4)}\right]\, ,
 \end{aligned}
 \end{equation}
 where $P_{s}$ and $P_{s}^{*}$ are given by
 \begin{equation}\label{Pconstants}
\begin{aligned}
P_{s}&=\frac{1}{2}+\frac{i s}{48 M \omega }\left(D_2 q_{s}- 2^s  q_{-s} C_{s}\mathcal{K}^2\right)\, ,\\
P_{s}^{*}&=\frac{1}{2}+\frac{i s}{48 q_{s}M \omega }\left(D_2-2^s C_{s} \mathcal{K}^2\right)\, ,
\end{aligned}
\end{equation}
$D_{2}$ is the Starobinsky-Teukolsky of the angular functions \req{D2value} and $\mathcal{K}^2=D_{2}^2+144 M^2 \omega^2$.
The origin of these $P_{s}$, $P_{s}^{*}$ constants is explained in the appendix. 

After removing the gauge freedom in the corrected Teukolsky equation by reducing the potential to its simplest form \req{eq:dVm2}, this now manifestly exhibits certain important properties.

\subsubsection*{Parity-preserving corrections}
First of all, the quasinormal mode frequencies should not depend not the ST constants, and at least in the case of parity-preserving corrections we can check this analytically. For those theories, the QNMs have definite parity and hence they are given by the polarization $q_{+2}=q_{-2}=\pm 1$. When we use those values of $q_{\pm 2}$ in \req{eq:dVm2}, using the coefficients $A_k$ computed for each theory, we observe that the dependence on $C_{s}$ drops off, leaving us with two different potentials $\delta V_{-2}^{\pm}$ --- one for each polarization --- that only depend on $M$, $a$ and $\omega$. Thus, the result is manifestly independent on the gauge choice for the metric reconstruction. Furthermore, the conjugate potential becomes identical $\delta V_{-2}^{*\pm}=\delta V_{-2}^{\pm}$, so that the full equations for $R_{-2}$ and $R_{-2}^{*}$ are identical and have the same solutions. This ensures the consistency of the construction. On the other hand, the potentials for $s=+2$,  that is, $\delta V_{+2}^{\pm}$,  also become independent of $C_{s}$ and they take a different form from $\delta V_{-2}^{\pm}$. However, they give rise to the same QNM frequencies, as checked in \cite{Cano:2023tmv} --- as remarked earlier the $s=+2$ and $s=-2$ variables contain the same information, so we can work with either of them. 

\subsubsection*{Parity-breaking corrections}
In the case of parity-breaking corrections we also expect the result to be independent of the choice of the $C_s$ constants, although in that case the equations have a more involved structure \cite{Cano:2023jbk}.
The key aspect about parity-breaking corrections is, however, that one has to determine the polarization $q_{\pm 2}$ by solving the different radial equations simultaneously. Since here we are only considering the $s=-2$ equations, we set $C_{-2}=0$, which has the effect of decoupling these from the $s=+2$ ones.

Then, we have to solve the $s=-2$ equation and its Newman-Penrose conjugate, which have different corrections to the potential, $\delta V_{-2}$ and $\delta V_{-2}^{*}$. Remarkably, we find that  these are given by\footnote{The denominators in these expressions are (up to a factor) the $P_{-2}$ and $P_{-2}^{*}$ constants \req{Pconstants}. }
\begin{equation}
\begin{aligned} 
\delta V_{-2}&=\frac{2}{1-i q_{-2} K }\delta V_{\rm break}\, ,\\
\delta V_{-2}^{*}&=-\frac{2}{1-i K/q_{-2} }\delta V_{\rm break}\, .
\end{aligned}
\end{equation}
for the same $\delta V_{\rm break}$ that is independent of $q_{-2}$, and where 
\begin{equation}
K= \frac{D_{2}}{12 M \omega}\, ,
\end{equation}
Now, since we are interested in finding quasinormal mode solutions, the equations for $R_{-2}$ and $R_{-2}^{*}$ must be satisfied at the same time for the same QNM frequency. Clearly, this will only happen if 
\begin{equation}
\frac{2}{1-i q_{-2} K }=-\frac{2}{1-i K/q_{-2} }\, ,
\end{equation}
so that the potentials become identical $\delta V_{-2}=\delta V_{-2}^{*}$.  This leads to the following solution for the polarization parameter
\begin{equation}
q_{-2}^{\pm}=\frac{i K}{1\pm \sqrt{K^2+1}}\, ,
\end{equation}
which, remarkably, is theory-independent. 
We also obtain the associated potentials for each of the polarizations
\begin{equation}
\delta V_{-2}^{\pm}=\delta V_{-2}^{*\pm}=\pm\frac{2}{\sqrt{1+K^2}}\delta V_{\rm break}\, .
\end{equation}
Thus, the polarizations $+$ and $-$ have opposite corrections to the QNM frequencies. This is a known property of parity-breaking higher-derivative corrections \cite{McManus:2019ulj,Cano:2021myl,Cano:2023jbk} and the fact that we can check it analytically provides a very strong test on the validity of our results.

\subsection{Convergence of the expansion}
In order to determine the radius of convergence of the spin expansion of the $A_k$ coefficients in \req{eq:Akcoefficients} we must study how the coefficients  $A_{k,n}$ behave for large $n$. Although we do not have a formula for $A_{k,n}$ for arbitrary $n$, our explicit results up to $n_{\rm max}=18$ provide a good intuition. First, let us note that these coefficients depend on the frequency, so in order to obtain some numerical values we can set $\omega$ to be close to the frequency of the QNM in which we are interested. Our results show very clearly that the coefficients $A_{k,n}$ grow exponentially with $n$, as
\begin{equation}\label{eq:Aknlargen}
A_{k,n}\sim c_k \Lambda_{1}^{n}\, ,
\end{equation}
for some complex number $\Lambda_{1}$ which depends on the frequency. Generically, we find that $|\Lambda_{1}|>1$, and hence the radius of convergence is $\chi^{\rm max}=|\Lambda_{1}|^{-1}<1$. However, this is just a problem of the series expansion, as we expect the full coefficients $A_{k}(\chi)$ to remain finite for every $\chi$. Indeed, what \req{eq:Aknlargen} suggests is that the full expression of $A_{k}$ can be written in the form
\begin{equation}\label{eq:Akpole}
A_{k}(\chi)=\frac{\tilde A_k(\chi)}{1-\Lambda_{1} \chi}\, .
\end{equation}
Here, the series expansion of the denominator precisely gives rise to the behavior \req{eq:Aknlargen}, but it does not introduce any divergence since $\Lambda_1$ is complex and $\chi$ is real. On the other hand, the series expansion of the numerator $\tilde A_{k}(\chi)$ should now have a much better convergence than the original $A_{k}(\chi)$, because we have absorbed the terms that spoiled the convergence in the denominator.

Now, the way to do this in practice is to transform our series expansion \req{eq:Akcoefficients} into a Pad\'e approximant with a denominator of order 1
\begin{equation}\label{PadeA}
A_{k}^{\text{Pad\'e}(1)}=\frac{1}{1-\tilde\Lambda_{1} \chi}\sum_{n=0}^{n_{\rm max}-1}\chi^n \tilde A^{(1)}_{k,n}\, .
\end{equation}
Remarkably, the value of $\tilde \Lambda_{1}$ obtained from the Pad\'e approximant coincides very well with the value of $\Lambda_{1}$ estimated from the asymptotic behavior of the coefficients. Thus, the Pad\'e approximant is very powerful in identifying the complex pole in \req{eq:Akpole}.  The new series expansion in the numerator of \req{PadeA} always has a better convergence than the original series \req{eq:Akcoefficients}, and there are two possibilities. If the coefficients $A^{(1)}_{k,n}$ remain bounded or grow slow enough for $n\rightarrow \infty$, then the  Pad\'e approximant \req{PadeA} has a radius of convergence of $\chi^{\rm max}=1$, and therefore it is convergent for all subextremal values of the spin. Thus taking $n_{\rm max}$ to be large enough we should be able to consider arbitrarily high spins. 
The second possibility is that the new coefficients still grow exponentially, $A^{(1)}_{k,n}\sim \Lambda_{2}^n$, with $1< |\Lambda_2|< |\Lambda_1|$. In that case, the Pad\'e approximant has a bigger radius of convergence than the original series, but it still diverges for high enough spin $\chi^{\rm max}=|\Lambda_2|^{-1}$. Thus, to improve the convergence we proceed in the same way and apply a Pad\'e approximant with order 1 denominator to the numerator of \req{PadeA}. Then we end up with an order-2 Pad\'e approximant, 
\begin{equation}\label{PadeA2}
A_{k}^{\text{Pad\'e}(2)}=\frac{1}{\left(1-\tilde\Lambda_{1} \chi\right)\left(1-\tilde\Lambda_{2} \chi\right)}\sum_{n=0}^{n_{\rm max}-2}\chi^n \tilde A^{(2)}_{k,n}\, ,
\end{equation}
where $\tilde\Lambda_{2}\approx \Lambda_{2}$ is again complex and hence the full expression is finite. We can then analyze the convergence of the new series expansion with coefficients $\tilde A^{(2)}_{k,n}$ and proceed in the same way. If the new series converges for all values of the spin then we are done, and otherwise we consider a Pad\'e approximant of one more order. 

In general, we get a Pad\'e approximant with an order-$N$ denominator
\begin{equation}\label{PadeAn}
A_{k}^{\text{Pad\'e}(N)}=\frac{\sum_{n=0}^{n_{\rm max}-N}\chi^n \tilde A^{(N)}_{k,n}}{\prod_{n=1}^{N}\left(1-\tilde\Lambda_{n} \chi\right)}\, .
\end{equation}
and we consider the minimum value of $N$ for which the numerator behaves as a convergent series $\forall |\chi|<1$. The $\tilde\Lambda_{n}$ are always complex and hence this expression is regular for real values of $\chi$. We note that increasing the value of $N$ beyond the minimum value required to achieve convergence does not always result in a faster convergence and it could in turn introduce spurious poles.

\begin{figure}[t!]
	\centering
	\includegraphics[width=0.48\textwidth]{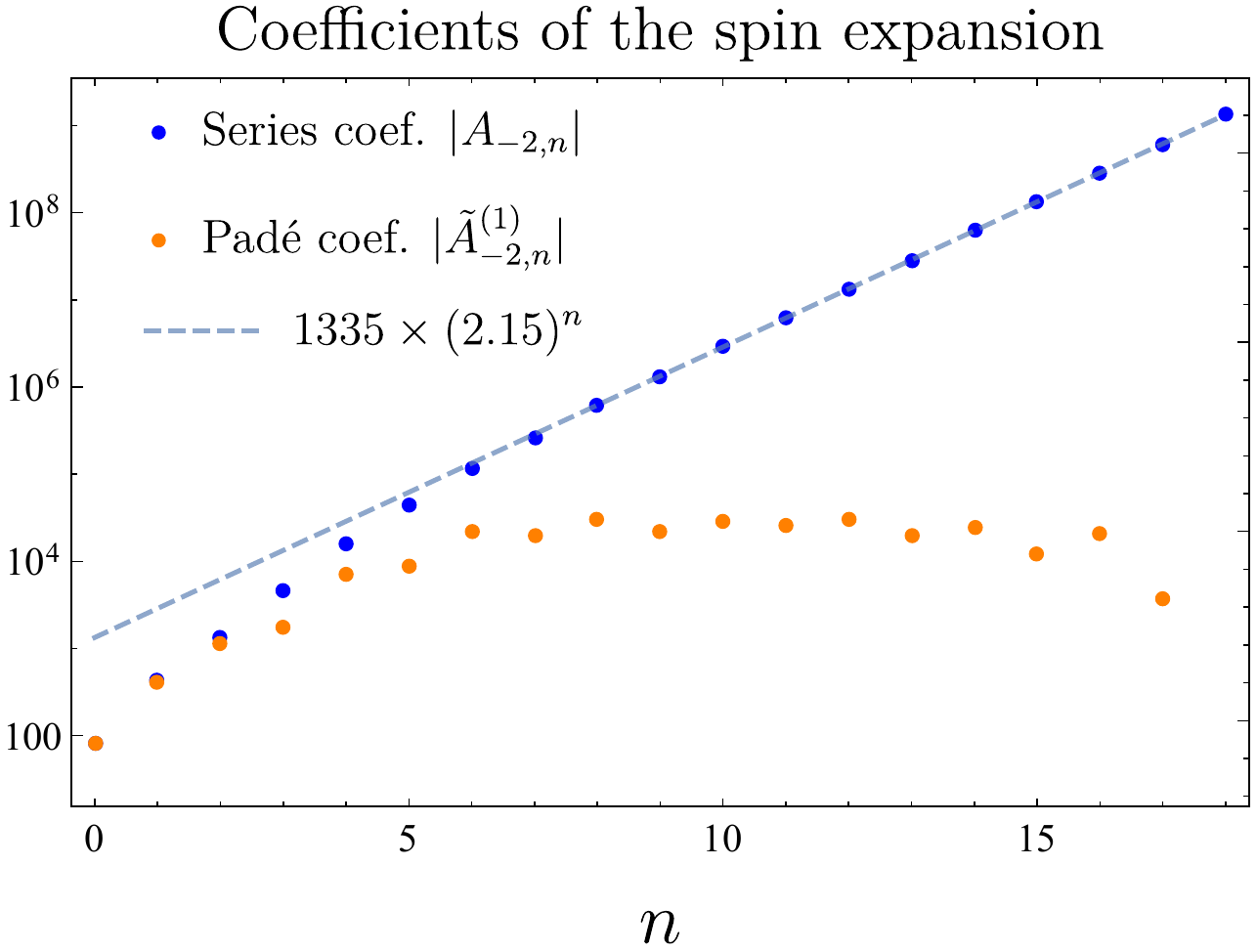}
	\caption{Absolute value of the coefficients $A_{-2,n}$ of the series expansion \req{eq:Akcoefficients} and $\tilde A_{-2,n}$ of the Pad\'e approximant \req{PadeA} as a function of $n$. We show the values corresponding to the $l=m=2$ mode with $q_{s}=-1$ of the quartic Lagrangian $\mathcal{C}^2$ and we set $M=1$ and $\omega=0.533 - 0.081$, corresponding to the $(2,2,0)$ QNM frequency of Kerr black holes with $\chi=0.7$. A fit reveals that the coefficients of the series expansion grow exponentially as in \req{eq:Aknlargen} with $|\Lambda_1|\sim 2.1547$. Thus, the radius of convergence is $\chi^{\rm max}=|\Lambda_1|^{-1}\sim0.46$. On the other hand, the coefficients of the Pad\'e approximant are bounded, indicating that the series in \req{PadeA} converges for all values of the spin $|\chi|<1$. The numerator of the Pad\'e approximant gives $\tilde \Lambda_1=0.2969 + 2.134 i$, whose absolute agrees perfectly with $|\Lambda_1|$.   }
	\label{fig:convergenceA}
\end{figure}

We have explored this strategy explicitly for the theories \req{eq:EFT} using an expansion of order $n_{\rm max}=18$ for the $l=2,3$ modes and of order $n_{\rm max}=14$ and for the $l=4$ ones. We have found that the order of the denominator of the Pad\'e approximant that we should use typically increases with the magnitude of the angular numbers $l$ and $|m|$. 
In practice, we used $N=0$ for  the $l=2,3$, $m=0$ modes, $N=1$ for $l=2$, $m\neq 0$, and $N=2$ for $l=3$, $m\neq 0$ and all the $l=4$ modes. 
Presumably, one needs to keep increasing the value of $N$ as $l$ and $|m|$ grow, but we do not have conclusive evidence of this. 
We illustrate the convergence of the Pad\'e approximant in Fig.~\ref{fig:convergenceA}, where we show the coefficients $A_{-2,n}$ and $\tilde A^{(1)}_{-2,n}$ of the $l=m=2$, $q_{s}=-1$ mode of the quartic theory $\mathcal{C}^2$.  Although these results are only based on numerical evidence, they strongly indicate that the Pad\'e approximant yields a convergent result. Therefore, this method allows us to study black holes of arbitrary spins as long as we include enough terms in the spin expansion.

\section{Quasinormal modes}\label{sec:QNMs}
Summarizing the results of the previous section, in order to study the quasinormal modes of the rotating black holes in the theory \req{eq:EFT}, it suffices to consider the equation \req{eq:correctedradial} with either $s=+2$ or $s=-2$, since both are equivalent. We focus on the $s=-2$ case, so we have to solve the equation
\begin{equation}\label{eq:correctedradial2} 
\begin{aligned}
\Delta^{3}\frac{d}{dr}\left[\Delta^{-1}\frac{dR_{-2}}{dr}\right]+\left(V_{-2}+ \alpha \delta V_{-2}\right) R_{-2}=&0\, ,
\end{aligned}
\end{equation}  
where $\delta V_{-2}$ is given by the simple expression \req{eq:dVm2}. This equation belongs to the class of parametrized modified Teukolsky equations analyzed in Ref.~\cite{Cano:2024jkd}, which obtains the corrections to the QNM frequencies using a continued fraction method.
Thus, the results of that work are directly applicable to study the QNMs of \req{eq:correctedradial2} and we refer to that paper for details. We note that a public \texttt{python} code for the computation of QNM frequencies is available on \texttt{github}~\cite{github}. 

Ref.~\cite{Cano:2024jkd} considers potentials of the form \req{eq:dVm2} with free coefficients $A_k$ (allowing for more values of $k$), but in our case these coefficients are not free but determined by the theory. Furthermore, they depend on the frequency $\omega$ that one is trying to determine. However, at first order in the higher-derivative couplings, it suffices to evaluate those coefficients on the corresponding Kerr QNM frequency whose corrections we want to obtain. Thus, for each value of harmonic numbers $(l,m)$, overtone index $n$, polarization $\pm$ and angular momentum $\chi$, each of the higher-derivative theories provides a numeric value for the coefficients $A_k$. Combining those values with the shift in the QNM frequencies associated to each individual $A_k$, computed in \cite{Cano:2024jkd}, we obtain the full shift in the QNM frequencies due to higher-derivative corrections. 

For each of the higher-derivative corrections, with couplings $\alpha_q$ defined in \req{alphaq}, we write
\begin{equation}\label{defdeltaomega}
\omega_{lmn}^{\pm}=\omega^{{\rm Kerr}}_{lmn}+\frac{\alpha_{\rm q}}{M}\delta\omega_{lmn}^{{\rm q},\pm}+\mathcal{O}(\alpha_{\rm q}^2)\,.
\end{equation}
Thus, we define $\delta\omega^{\rm q}$ to be the coefficients of the linear-order corrections to the QNM frequencies. We note that these quantities are defined in a way such that they are dimensionless and they only depend on the dimensionless spin, $\delta\omega^{\rm q}=\delta\omega^{\rm q}(\chi)$. Our perturbative analysis of higher-derivative terms only allows us to compute these linear coefficients, but it is also interesting to look at the non-linear terms in \req{defdeltaomega} in order to determine the regime of validity of the linear corrections. We come back to this later.

We note that these frequencies exhibit the same symmetries as in Kerr, and in particular considering negative values of $\chi$ is equivalent to the exchange $m\rightarrow -m$:
\begin{equation}
\delta\omega_{lmn}^{{\rm q},\pm}(-\chi)=\delta\omega_{l-mn}^{{\rm q},\pm}(\chi)\, .
\end{equation} 

By using the results of \cite{Cano:2024jkd}, we have obtained numerically the values of $\delta\omega_{lmn}^{{\rm q},\pm}$ for all the higher-derivative theories in \req{eq:EFT}, for $l=2,3,4$, $m\in [-l,l]$, $n=0,1,2$ and $0\le \chi\le 0.95$ in the case of $l=2,3$ and $0\le \chi\le 0.75$ for $l=4$. 
Due to the large amount of data, we provide these results in the form of .txt files in the \texttt{github} repository \cite{gitbeyondkerr}. These files contain, for a list of values of the spin, the real and imaginary part of $\delta\omega^{\rm q}$ for the corresponding mode and theory. They also contain the error in both the real and imaginary parts, estimated from the convergence of the spin expansion. Namely, the error is the difference between the predictions with spin expansions of order $n_{\rm max}$ and of order $n_{\rm max}-1$. In this case, $n_{\rm max}=18$ for the $l=2,3$ modes and $n_{\rm max}=14$ for the $l=4$ ones. 

\begin{figure*}
	\centering
	\includegraphics[width=0.98\textwidth]{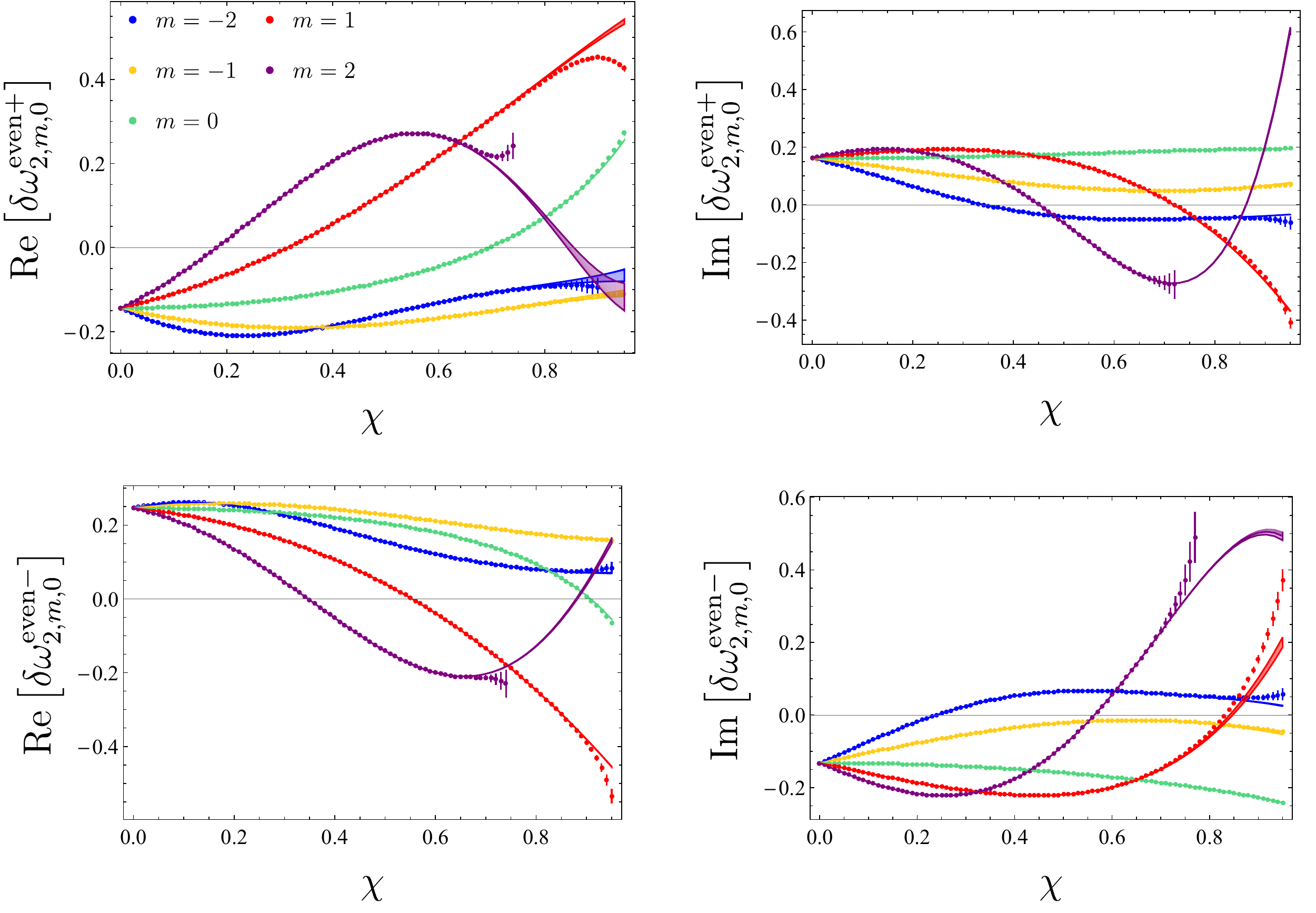}
	\caption{Shifts in the QNM frequencies of the $l=2$ fundamental modes due to the even-parity cubic curvature corrections. The dots represent data points and the error bars are estimated by looking at the convergence of the spin expansion --- they correspond to the difference with respect to the prediction obtained with one less order. The lines and shaded regions correspond to the fitting polynomials of order $N_0-1$, $N_0$ and $N_0+1$, where $N_0$ is the optimum order.}
	\label{fig:evenl2}
\end{figure*}

We observe that the convergence of the spin expansions depends a great deal on the value of $m$. Typically, the $m=l$ modes have the slowest convergence, and we get reliable results up to $\chi\sim 0.7-0.8$. For smaller values of $m$, including negative values, the convergence is much faster, and for some modes we get an accurate result up to the maximum value of the spin that we computed, $\chi\sim 0.95$. As an example, in Fig.~\ref{fig:evenl2} we offer a visualization of the $\delta\omega^{\rm q}$ coefficients of the $(l,m,n)=(2,m,0)$ modes of the even-parity cubic theory.

As a test of our results, we have compared them with those of \cite{Cano:2023jbk} for the $(l,m,n)=(2,2,0), (3,3,0)$ modes, which were obtained using an eigenvalue perturbation approach to solve the modified Teukolsky equation \cite{Zimmerman:2014aha,Hussain:2022ins}.  We find that the two results agree quite well, with a relative difference in $\delta \omega^{\rm q}$ that increases approximately linearly with $\chi$, and which is smaller than $9\%$ for $\chi<0.7$. However, we found that the eigenvalue perturbation method that was employed in \cite{Cano:2023jbk} was incomplete, as it did not take into account the variation in the angular separation constants $B_{lm}$ due to the correction to the frequency. This explains the small difference between our results here and those in \cite{Cano:2023jbk}. In fact, we have repeated the eigenvalue perturbation computation of $\delta\omega^{\rm q}$ including this effect, finding that the relative difference between that method and Leaver's method for $|\chi|\le 0.4$ is $\lesssim 10^{-6}$ if $l =2$ and $\lesssim 10^{-5}$ if $l=3$.  We have checked that this is the case as well for overtones, which were not computed in \cite{Cano:2023jbk}. This impressive agreement serves as a consistency test of the methods and results we obtain here.

\subsection{Polynomial fits}
Although we provided numerical results for the shifts in the QNM frequencies, it is most useful to express those results in terms of a fitting function that captures the dependence on the angular momentum. Here we follow a strategy similar to that of \cite{Cano:2023jbk} and write our results in terms of polynomial functions using a weighted fit. 

We consider the data set $\{(\chi_i,\delta\omega_i),i=1,\ldots, i_{\rm max}\}$, where $\delta\omega_i$ are the numerically computed values of $\delta\omega_{lmn}^{{\rm q},\pm}$ (we are dropping the labels for clarity) for $\chi=\chi_i$, and we include both positive and negative values of $\chi_i\in [-\chi_{\rm max},\chi_{\rm max}]$, with $\chi_{\rm max}=0.95$ for $l=2,3$, and $\chi_{\rm max}=0.75$ for $l=4$. We perform independent fitting procedures for the real and imaginary parts of the frequency, but in order to avoid introducing more notation let us imagine that $\delta\omega_i$ denotes either the real or the imaginary part. 

The data points $\delta\omega_i$ have two sources of error. On the one hand, there is an error coming from the tolerance of the numerical method employed for the computation of QNM frequencies, which is around $\varepsilon_{\rm num}\sim 10^{-7}$ \cite{Cano:2024jkd}. On the other hand, there is an error due to the convergence of the spin expansion, and we estimate it as the difference of the results using spin expansions of order $n_{\rm max}$ and of order $n_{\rm max}-1$, as we mentioned above. Let us denote this error by $\varepsilon_{{\rm spin}, i}$, which indeed depends on the spin and it is different for each data point. We only include data points for which $\varepsilon_{{\rm spin}, i}$ is less than $10\%$ of the maximum value of $|\delta\omega_i|$ $\forall i$.
Thus, the total error is 
\begin{equation}
\varepsilon_{{\rm total},i}=\varepsilon_{\rm num}+\varepsilon_{{\rm spin},i}\, .
\end{equation}
For small $\chi$ the spin expansion is very accurate and we have $\varepsilon_{{\rm spin},i}\ll \varepsilon_{\rm num}$, so the error from the numerical method is the bottleneck. For large spins, the error from the spin expansion dominates, with the transition happening at some critical value $\chi_c$ which becomes larger as we increase the order of the expansion. 

Then, we wish to fit our data points to an $N$th-degree polynomial
\begin{equation}\label{eq:fitdeltaw}
\delta\omega_{\rm fit}^{N}(\chi)=\sum_{j=0}^{N}c_{j}\chi^j\, .
\end{equation}
In order to take the different accuracy of each data point into account, the fits are obtained by minimizing the weighted sum of squared residuals
\begin{equation}
\Sigma_{N}=\sum_{i=1}^{i_{\rm max}}\frac{1}{\varepsilon_{{\rm total},i}^2}\left|\delta\omega_i-\delta\omega_{\rm fit}^{N}(\chi_i)\right|^2\, ,
\end{equation}
where each weight is the inverse of the estimated variance of the corresponding data point. 
In order to determine the optimum degree of the fitting polynomial, we compute the two-step ratios of the minimized error functions
\begin{equation}
d_{N}=\frac{\Sigma_{N+1}^{\rm min}}{\Sigma_{N-1}^{\rm min}}\, ,
\end{equation}
with $N=1,2,\ldots$. For the first few values of $N$, we typically have $d_N\ll 1$, indicating that increasing the degree of the polynomial results in a large increase of the goodness of the fit. For large enough $N$, we find that $d_N$ saturates and approaches one, so increasing $N$ does not significantly improve the fit. Thus, in order to avoid overfitting the data, we choose a value of $N$ such that $d_N$ is about to saturate. This is the maximum degree of the polynomial that guarantees that the fit is reliable and that we do not introduce spurious behavior due to overfitting. For definiteness, we defined this optimum degree $N_0$ to be the lowest value of $N\ge 2$\footnote{The reason to start at $N=2$ is that, in rare cases, the value of $d_N$ is not small for the first few values of $N$, but it does decrease after that.} for which $d_{N_0}>0.2$.    

We find that $N_0$ varies a lot: depending on the mode and the theory, we can have $6\le N_0\le 18$. The reason for this is not only the different accuracy of the results in each case, but it also has to do with the variability of $\delta\omega$ as a function of the spin. For some modes, $\delta\omega(\chi)$ has a mild dependence on $\chi$ and a low-order polynomial can fit it with very high accuracy. For others, it has a great variability (\textit{e.g.}, it grows very fast, or shows several maxima and minima), so that a higher-degree polynomial is needed. We remark in particular that the order of the polynomial need not coincide with the order of the spin expansion, since the fitting polynomials are not the same as a Taylor expansion. We provide the coefficients of the optimum fitting polynomials in the \texttt{github} repository \cite{gitbeyondkerr}.

We expect that these fitting polynomials provide a reliable estimation for $\delta\omega$ even beyond the fitting range of $\chi$. We show a test of this in Fig.~\ref{fig:evenl2}. Along with the numerical data points, this figure shows the corresponding polynomial fits of degree $N_0$ and $N_0\pm 1$. The shaded regions represent the area spanned by these three different polynomials, and they serve as a measure of the uncertainty in their prediction. We see that, even for the most problematic, $l=m=2$ modes, the three fits agree very well, even beyond the fitting range. Thus, it is conceivable that the prediction from these fits is relatively accurate for spins as high as $\chi\sim 0.9$. For the other modes, where we have a better convergence of the spin expansion, the result is even more robust.
It would be very interesting to check the validity of these results at high spins, but the methodology to compute the QNMs of highly spinning black holes in beyond-GR theories has not net been developed yet --- the results of \cite{Cano:2024bhh} may provide a promising direction for this though.  

In the fitting region, all these polynomials are virtually indistinguishable from the data points, except when the error bars become large. In some cases, the fit departs from those data points, but this is the way in which the fit tells us that those points are not credible.

\subsection{Overtones}
Our results include, for the first time, the corrections to the overtones of rotating black holes. Thus, we take the chance to discuss some of the properties of these overtones. 

As noticed long ago \cite{Nollert:1996rf,Nollert:1998ys}, and recently revived by \cite{Jaramillo:2020tuu,Jaramillo:2021tmt}, the spectrum of quasinormal modes of a black hole is unstable under small changes of the non-hermitian operator that defines the problem. The overtones are especially sensitive to this instability, and even a small deformation of the equations governing black hole perturbations can lead to a total disruption of the spectrum of overtones. 
Thus, the corrections to the overtones are expected to be much larger than the corrections to the fundamental modes.\footnote{Here we are excluding ``artificial'' modifications of the perturbation equations, such as those involving adding an extra small bump to the potential at a large radius \cite{Cheung:2021bol}, which are known to disrupt the full spectrum of QNMs. The modifications  due to higher-derivative corrections preserve the qualitative form of the potential and hence do not cause such a dramatic effect on the QNMs.} This effect has already been observed in some specific cases, see  \cite{Volkel:2022aca,Konoplya:2022iyn,Konoplya:2022pbc,Silva:2024ffz}.

\begin{figure}[t!]
	\centering
	\includegraphics[width=0.48\textwidth]{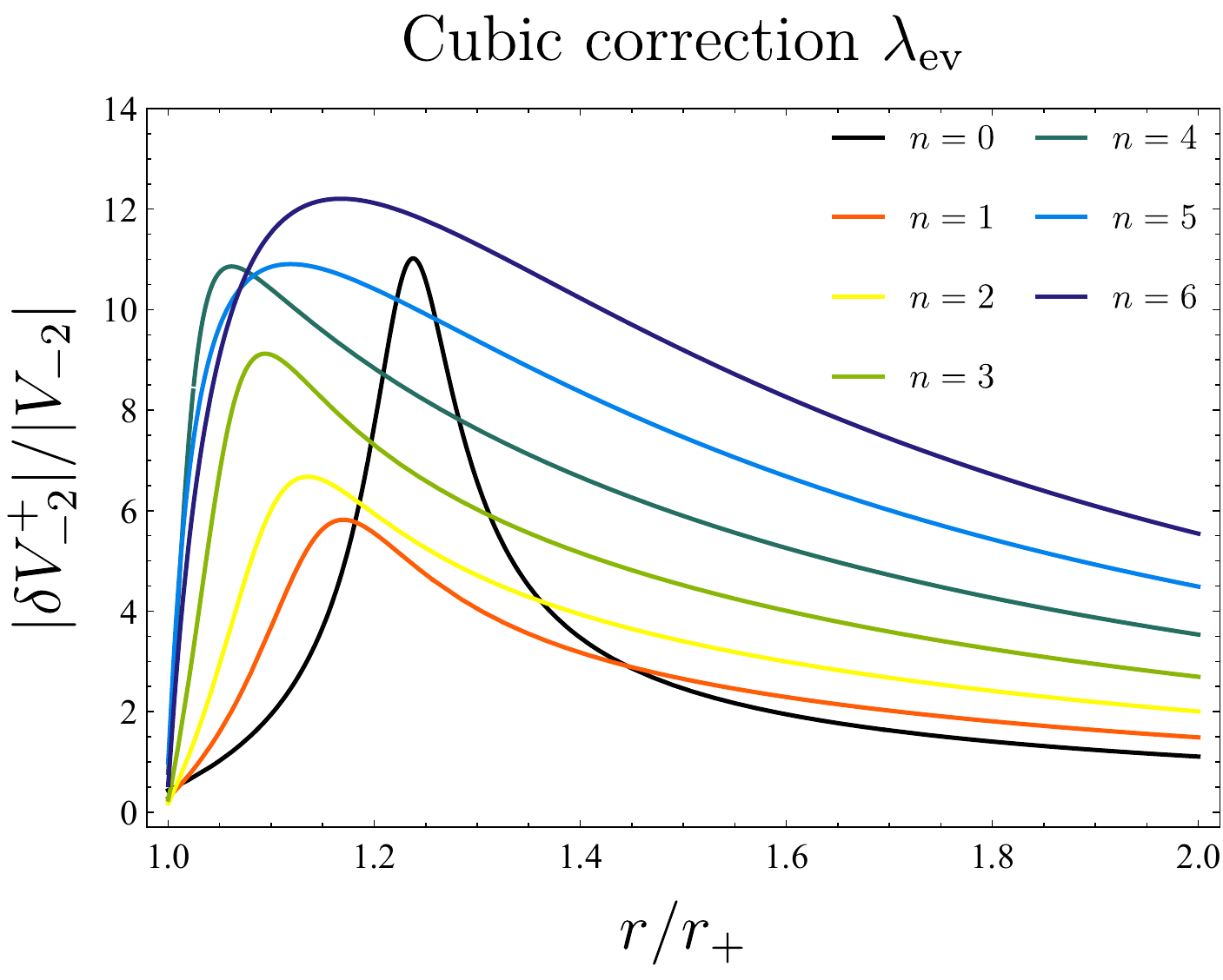}
 \includegraphics[width=0.48\textwidth]{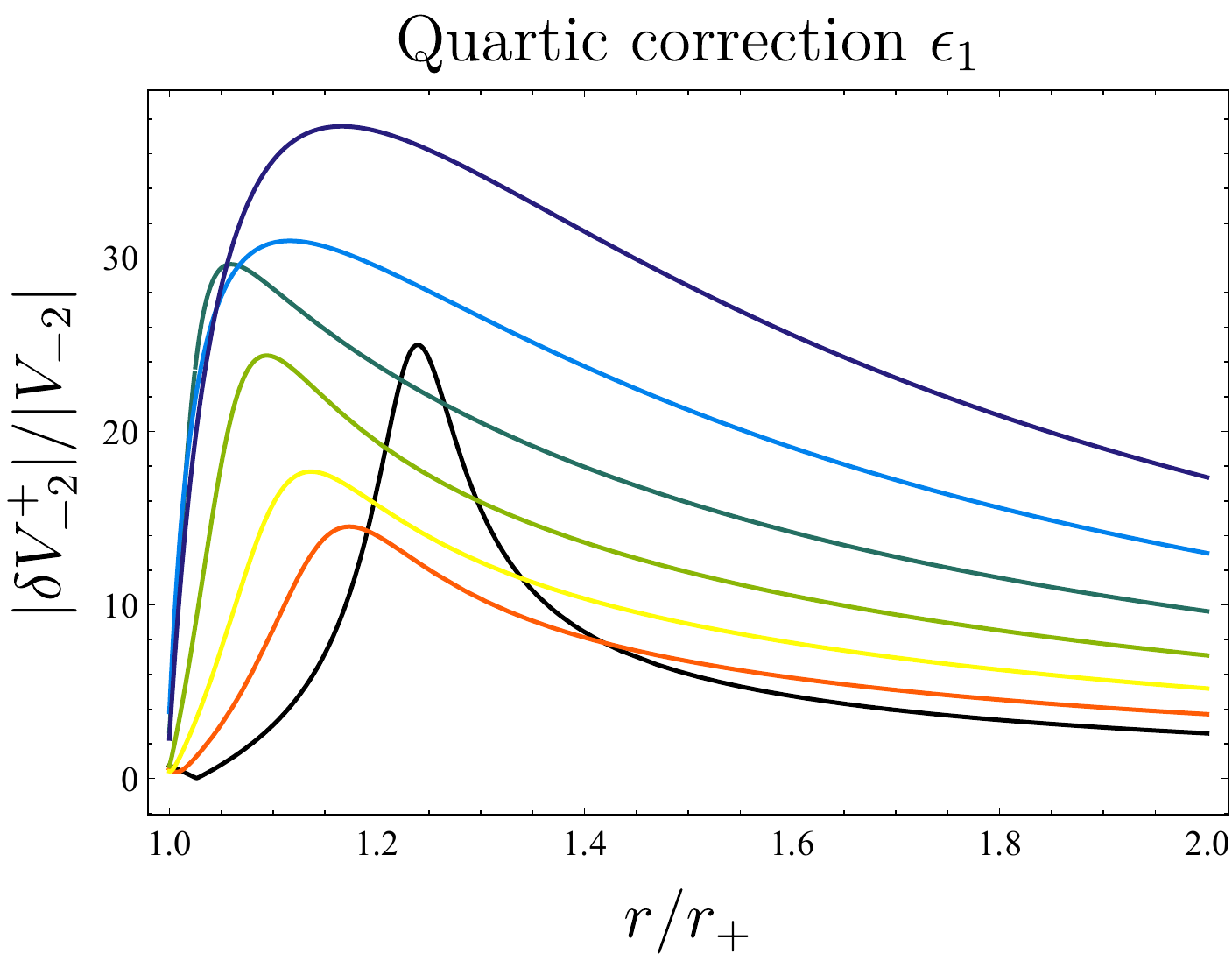}
	\caption{Relative correction to the potential for the first few overtones of modes with $l=m=2$ and ``$+$'' polarization for a black hole of spin $\chi=0.3$. The top plot corresponds to the even-parity cubic theory, and the bottom plot to the quartic theory labelled by the parameter $\epsilon_1$.}
	\label{fig:dV}
\end{figure}

In the case of higher-derivative corrections, there is yet another mechanism that will cause the corrections to the overtones to grow: the breakdown of the EFT at large frequencies. Indeed, for overtones of Schwarzschild and Kerr black holes, the $n$th-overtone frequency $\omega_n$ grows linearly with $n$ in the imaginary direction \cite{Berti:2009kk}. Now, when $|\omega|\rightarrow\infty$, the potential $V_s$ in the Teukolsky equation grows as $V_s\sim \omega^2$, but the correction to the potential $\delta V_{s}$ can typically grow with a higher power of $\omega$, hence producing larger corrections for higher overtones \cite{Silva:2024ffz}. 
However, a direct comparison of $\delta V_{s}$ with $V_{s}$ is subtle (especially in the rotating case) since the powers of $\omega$ appear multiplying different powers of $r$.  By exploring a few cases, we have found that the relative correction $|\delta V_{s}|/|V_s|$ indeed tends to grow for large enough $n$, although it does not necessarily show a monotonic behavior for the first few values of $n$, since the growth of $|\omega_n|$ with $n$ is relatively slow. We illustrate this in Fig.~\ref{fig:dV}, where we see that for the first few values of $n$ the correction to the potential actually decreases, although it eventually grows again. The growth appears to be faster in the quartic theory than in the cubic theory. On the other hand, the maximum of $|\delta V_{s}|/|V_s|$ tends to be closer to the horizon for higher overtones, although this behavior does not seem to be  universal.

\begin{figure*}
	\centering
	\includegraphics[width=0.98\textwidth]{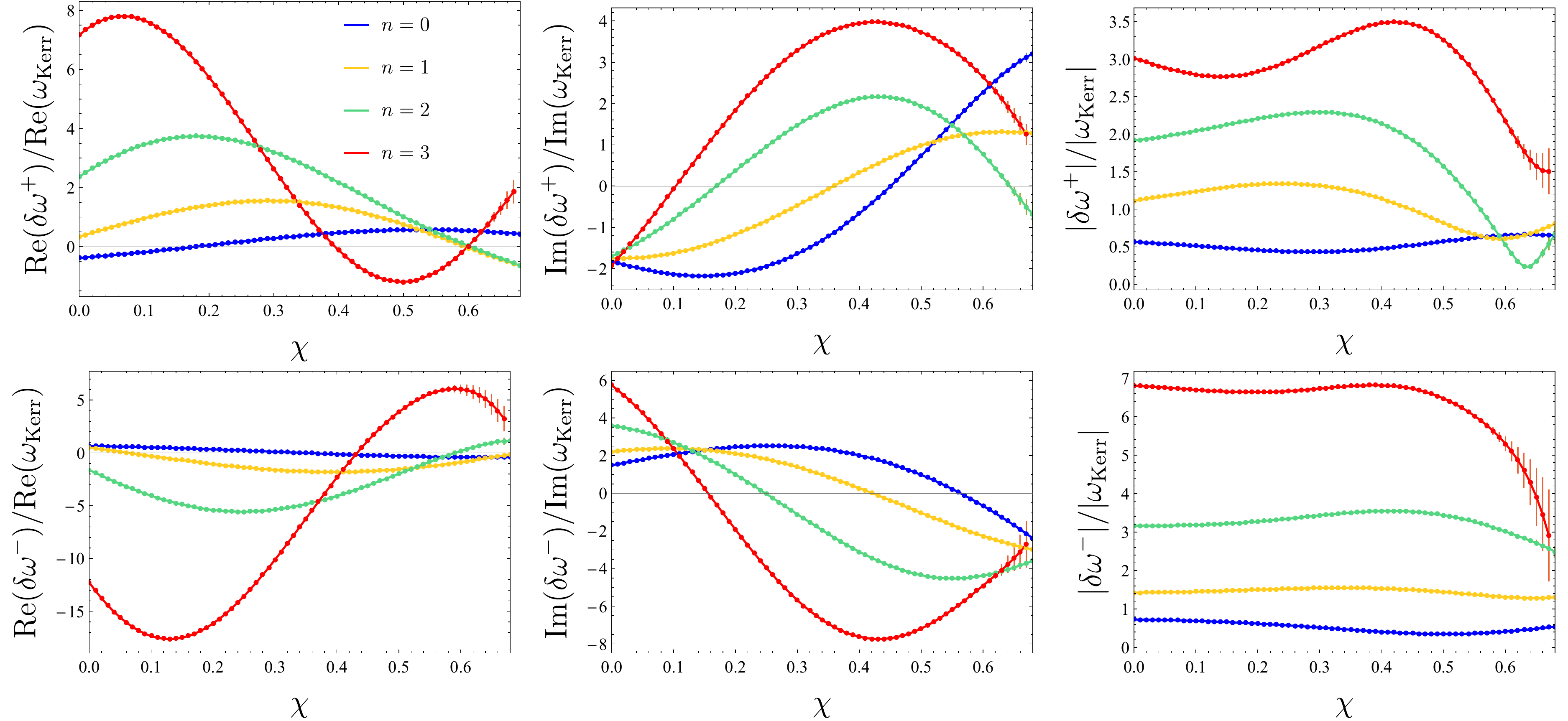}
	\caption{Corrections to the fundamental QNM frequency and to the first three overtones of $l=m=2$ modes in the even-parity cubic curvature theory. The top and bottom row correspond to $+$ and $-$ polarization, respectively. The first column shows the relative corrections to the real part of the frequency, the middle column the relative corrections to the imaginary part, and the last column, the absolute value of the relative correction.}
	\label{fig:evenovertones}
\end{figure*}
Let us then show the corrections to the QNM frequencies. For definiteness we discuss the case of the even-parity cubic theory, but our full results are available in the \texttt{github} repository \cite{gitbeyondkerr}. Fig.~\ref{fig:evenovertones} shows the relative corrections to the QNM frequencies of the fundamental mode and three first overtones with $l=m=2$ and for each of the two polarizations. The third column in that figure shows the relative correction in absolute value, and it reveals that, quite consistently, higher overtones receive larger corrections. We recall from Fig.~\ref{fig:dV} that the corrections to the potential for these low overtones is not necessarily bigger than the correction to the potential of the fundamental mode. Thus, this growth is probably related to the own unstable nature of the overtone spectrum. Although we have not computed higher overtones, we expect that these receive even larger corrections. 
The left and middle columns Fig.~\ref{fig:evenovertones} also offer interesting information: the corrections to the real and imaginary parts of overtones vary a lot depending on the angular momentum of the black hole, with the curves becoming more nonlinear as we increase the overtone index.  
It would be interesting to understand how these overtones behave for angular momentum close to extremality, but a rigorous analysis of this limit is not possible with our current methods. In fact, we observe that the spin expansion converges more slowly as we increase the overtone index.  The analysis of highly damped modes of rotating black holes thus requires an independent analysis.  Our results with $n=1,2$ are nevertheless still accurate in the vicinity of $\chi\sim 0.7$ and beyond depending on the case.

\subsection{Regime of validity of the linear approximation}
As we mentioned earlier, our analysis of the higher-derivative corrections is limited to first order in the coupling constants, which remains valid as long as these couplings are small enough. However, it is important to estimate exactly how small these couplings need to be. On the one hand, the modified Teukolsky equation  receives corrections with higher powers of the $\alpha_{\rm q}$ couplings, but these are out of reach for now.  On the other hand, even if we just consider the linearly modified Teukolsky equation, the QNM frequencies are in general non-linearly corrected. Here we explore these non-linear corrections to the QNM frequencies in order to estimate the regime of validity of the linear expansion, although we remark that this analysis is incomplete as we are missing the non-linear corrections to the Teukolsky equation. 

In our previous work \cite{Cano:2024jkd}, we assessed the reliability of the linear approximation for single $A_{k}$ modifications in the potential of the Teukolsky equation. Here we perform a similar analysis for the specific case of higher-derivative gravity, namely considering the potential modification of Eq. \req{eq:dVm2}. We want to compare the prediction obtained with the linearized framework with the one from the full continued fraction method. In order to do so, we consider the following notion of error, 
\begin{equation}
    \Delta_{\delta\omega}\equiv \left|\frac{\omega_{\text{nl}}-\omega_\text{lin}}{\omega_\text{lin}-\omega_{\text {Kerr}}}\right|\,.
\end{equation}
where $\omega_\text{lin}$ is the frequency \req{defdeltaomega} truncated at first order in $\alpha_{\rm q}$ and  $\omega_{\text{nl}}$ represents the exact non-linear result obtained from the continued fraction method, whose details of implementation can be found in~\cite{Cano:2024jkd}. 
This definition can be viewed as a relative error \emph{on the correction} $\delta \omega$ to the GR quasinormal frequencies. We remark that this is different from the relative error on the full frequency, as $\delta \omega$ is itself small.  Thus, the linear approximation remains valid as long as $\Delta_{\delta\omega}\ll 1$.

\begin{figure}[t!]
	\centering
	\includegraphics[width=0.44\textwidth]{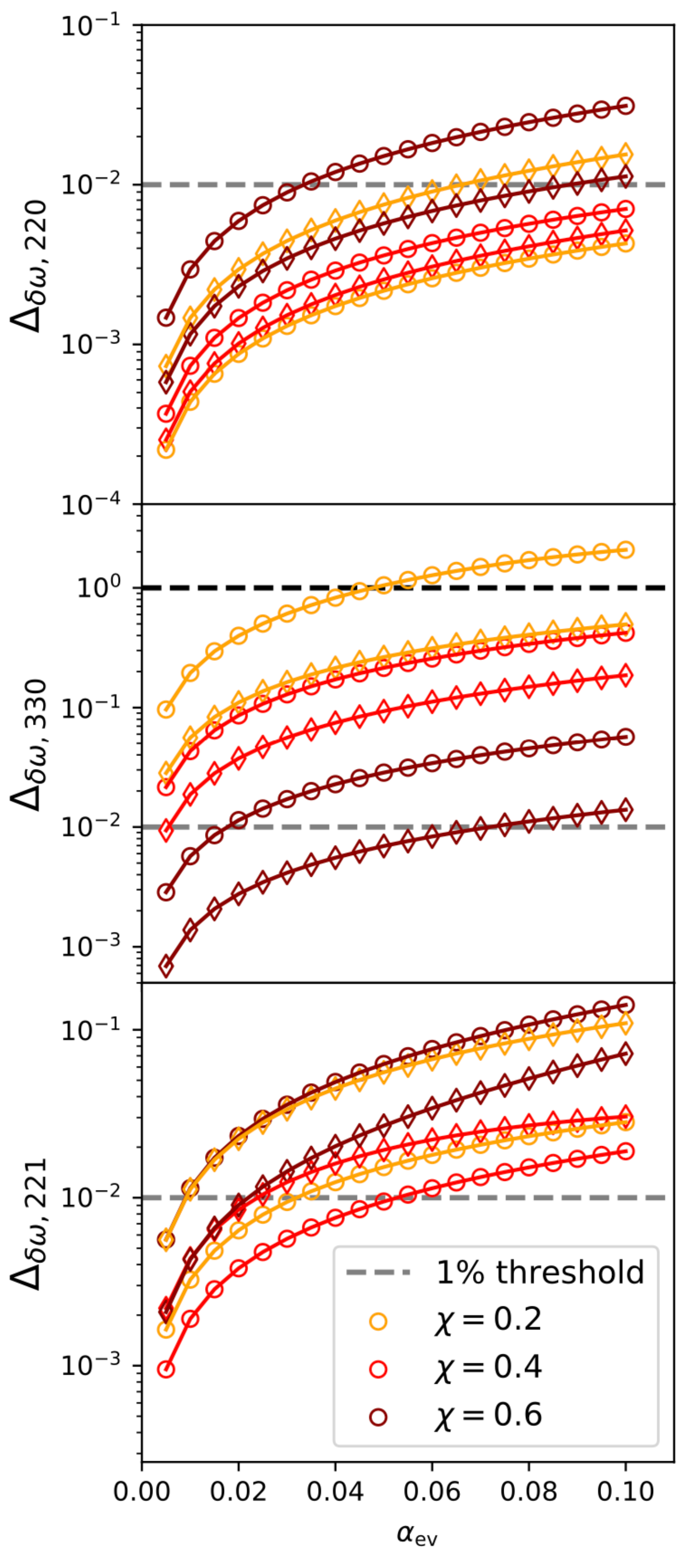}
	\caption{The relative error on the linear frequency correction versus dimensionless coupling in the cubic even theory for the $220$  (upper panel), $330$  (middle panel) and $221$ (lower panel) modes at different spin parameters. The circles correspond to plus polarization and the squares correspond to minus polarization. The threshold at which the error reaches $1\%$ and $100\%$ values are indicated with dashed lines.}
	\label{fig:ErrHDG}
\end{figure}

\begin{figure}
	\centering
	\includegraphics[width=0.47\textwidth]{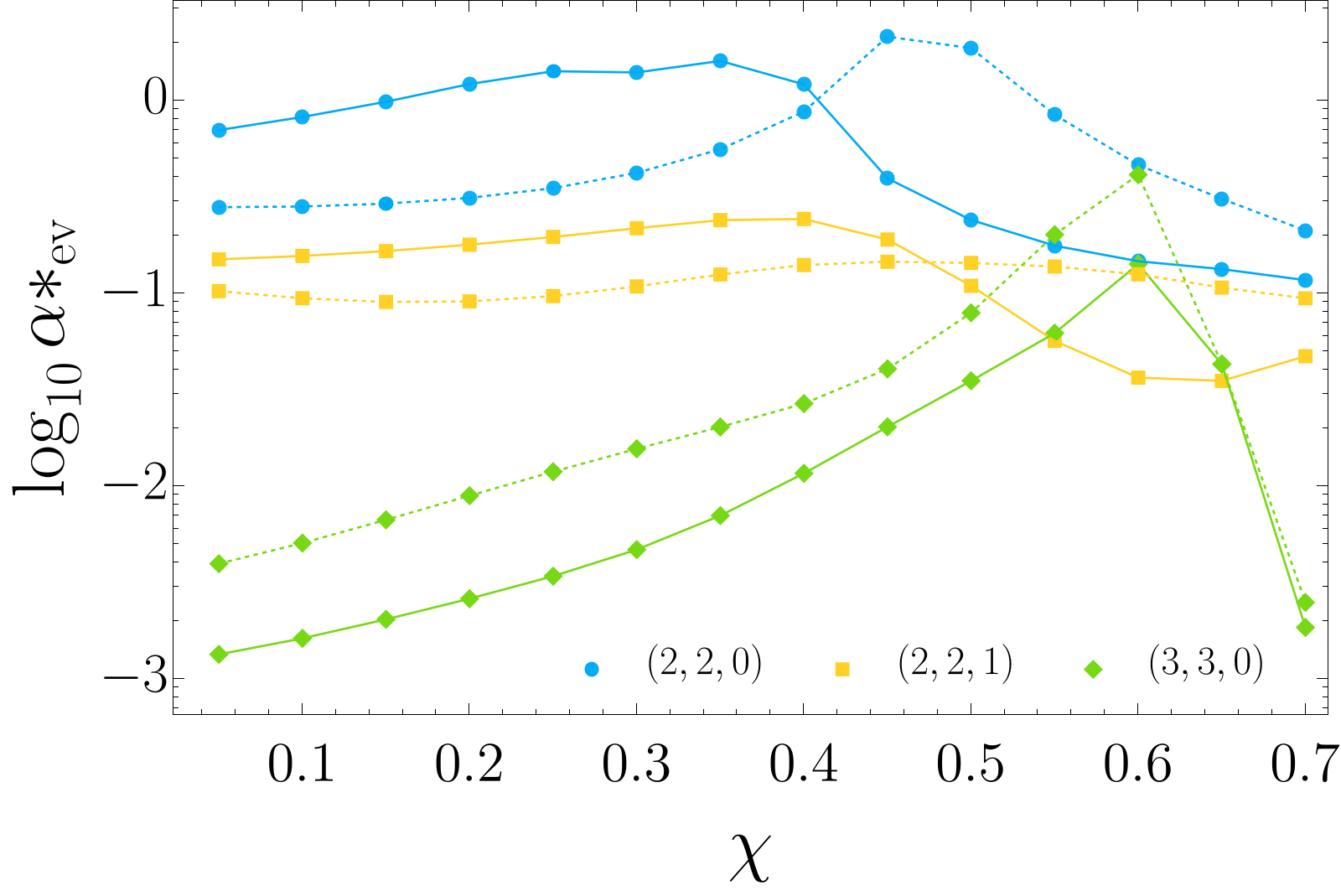}
	\caption{Threshold value of $\alpha_{\rm ev}^{*}$, for which the error between linear and non linear approximation becomes larger than $5\%$ as a function of the spin. We report $\alpha_{\rm ev}^{*}$ for the $(2,2,0)$, $(2,2,1)$ and $(3,3,0)$ modes (respectively circle, square and diamonds) for the "$+$" (solid lines) and "$-$" (dashed lines) polarizations of the even-parity cubic curvature theory.}
	\label{fig:ErrHDG5percent}
\end{figure}

In Fig.~\ref{fig:ErrHDG} we show the results for the $220$, $330$ and $221$ modes of the even-parity cubic theory at different values of the spin parameter for a range of dimensionless couplings $\alpha_{\rm ev}=\ell^4\lambda_{\rm ev}/M^4$. As we naturally expect, these plots show that the error of the linear approximation grows as the value of coupling increases. In Fig.~\ref{fig:ErrHDG5percent} we report the value of the coupling $\alpha^{*}_{\rm ev}$ for which $\Delta_{\delta\omega} = 5\%$ for a given mode at a given spin. We see that for $220$ this happens for $\alpha_{\rm ev}\sim 0.1-1$, which are actually very large values of the coupling. For reference, at $\alpha_{\rm ev}\sim 0.1$ the Kerr QNM frequencies receive corrections of the order of $20\%$, and realistically we do not expect the EFT corrections to be valid beyond that. Thus, the linear approximation seems to perform better than expected.  For the first overtone $221$ the breakdown of the linear approximation occurs for somewhat smaller (but still fairly large) couplings, which is consistent with the fact that overtones receive larger corrections and with the observations of \cite{Silva:2024ffz}.

On the other hand, we observe that for the $330$ modes, the linear approximation breaks down at much smaller couplings, especially for small spin. In this case we suspect that this is an artifact of the missing non-linear terms in the Teukolsky equation. Let us explain why. In the modified Teukolsky equation \req{eq:correctedradial2}, the correction to the potential is only defined up to terms that do not modify the frequencies at linear order in the coupling. Let us denote such kind of term by $\delta V_{\rm trivial}$. Thus, if one makes $\delta V\rightarrow \delta V+C \delta V_{\rm trivial}$, where $C$ is a constant, this has no effect at linear level in the coupling. However, $\delta V_{\rm trivial}$ will in general affect the frequencies non-linearly in the coupling, and this effect can be arbitrarily large if we take $C$ to be large. This is what seems to be happening for the $330$ modes.  
We observe that for these modes the $A_k$ coefficients in the Teukolsky equation are two orders of magnitude larger than for the $220$ and $221$ modes, but the linear correction to the frequency is of the same order of magnitude for all these cases.  Thus, the potential for the $330$ mode likely contains a large ``trivial'' piece $\delta V_{\rm trivial}$, which was probably introduced by all the manipulations we did to rewrite the potential in the form \req{eq:dVm2}.  This ambiguous term can only be fixed by computing the corrections to the Teukolsky equation at higher orders in the coupling. 
The conclusion is that our results should only be taken as a preliminary exploration, as the missing terms in the Teukolsky equation are crucial.  

In any case, our results do confirm that a linear correction to the Teukolsky equation does imply a linear correction to the frequencies within a reasonable range of couplings. 
Another interesting observation that follows from Fig.~\ref{fig:ErrHDG5percent} is that, contrarily to what one could naively expect, increasing the rotation does not seem to imply an earlier breakdown of the linear regime.

\section{Conclusions}\label{sec:conclusions}
We have provided the most complete calculation up to date of the corrections to the QNMs of Kerr black holes in a general EFT extension of GR. All our results can be found in the repository \cite{gitbeyondkerr} that contains the analytically computed corrections to the potential \req{eq:dVm2}, the numerical values of the shift in the QNM frequencies as defined in \req{defdeltaomega} and the coefficients of the polynomial fits \req{eq:fitdeltaw}. We computed these shifts for all the modes $(l,m,n)$ with $l=2,3,4$, $-l\le m\le l$, $n=0,1,2$ of both polarizations $\pm$ for all the theories in \req{eq:EFT} and for a large range of the black hole spin $\chi$. The maximum spin that we can accurately capture depends on the mode, and it ranges between $\chi_{\rm max}\sim 0.7$ and $\chi_{\rm max}\sim 0.95$.  We also performed a preliminary analysis of non-linear corrections to the QNM frequencies, finding that the first-order correction is accurate within a reasonable range of coupling constant values.  Thus, these results should allow us to perform complete black hole spectroscopy tests of higher-derivative gravity with ringdown observations. 

Although our results capture the modes and black hole solutions that are most relevant for astrophysical observations, there are still many unanswered questions about the QNM spectrum of rotating black holes beyond GR. For instance, our analysis does not include yet highly damped modes or eikonal modes, since they require a separate treatment. Eikonal modes are especially interesting due to their correspondence with unstable null geodesics in GR \cite{Cardoso:2008bp,Yang:2012he}, which must have a generalization in the case of higher-derivative gravity \cite{Bryant:2021xdh,Cano:2024wzo}.

More importantly, the QNMs of highly rotating (\textit{i.e.}, near-extremal) black holes cannot be accessed with our current methods based on a series expansion in the spin. The case of near-extremal black holes is particularly interesting since new phenomena appears close to extremality \cite{Yang:2012pj} and a possible amplification of higher-derivative effects could take place \cite{Horowitz:2023xyl,Horowitz:2024dch,Chen:2024sgx}. 
New methods will be required to study the QNM spectrum of these black holes. The analysis of perturbations in the near-horizon region recently reported by \cite{Cano:2024bhh} may provide a promising starting point.  

In addition, our analysis of the non-linear behavior in the coupling was incomplete as we are missing the second-order corrections to the Teukolsky equation. Doing this would require being able to reconstruct the metric perturbation at first order in the coupling, which is challenging. As a more accessible case, one could consider static or slowly rotating black holes, whose perturbations can be studied explicitly in terms of the metric \cite{Cardoso:2018ptl,Cano:2021myl,Pierini:2022eim,Franchini:2023xhd}, or with the new approach of \cite{Mukkamala:2024dxf}. In those cases it would be possible to find the corrections to the Regge-Wheeler and Zerilli equations at second order in the coupling, and this would allow us to rigorously estimate the regime of validity of the linear corrections to the QNM frequencies.

Finally, it would be interesting to extend our analysis to the case of theories with scalar fields such as dynamical Chern-Simons gravity \cite{Campbell:1990fu,Alexander:2009tp}, scalar-Gauss-Bonnet gravity \cite{Kanti:1995vq} and string gravity \cite{Cano:2021rey}. The modified Teukolsky formalism is also applicable to those cases \cite{Li:2022pcy,Hussain:2022ins}, yielding coupled equations for scalar and gravitational perturbations. However, these equations have so far only been obtained to first order in the angular momentum for dynamical Chern-Simons gravity \cite{Wagle:2023fwl}, so a more general analysis would be required. In the case of dilaton-Gauss-Bonnet gravity, the results from the Teukolsky analysis could then be contrasted with recent computations from spectral methods \cite{Chung:2024ira,Chung:2024vaf,Blazquez-Salcedo:2024oek}.
We leave all these questions for future work.

\begin{acknowledgments} 
P.~A.~C.~and~S.~M.~would like to thank the Strong group at the Niels Bohr Institute (NBI) for their kind hospitality during the final stages of this work. The work of P.~A.~C.~received the support of a fellowship from “la Caixa” Foundation (ID 100010434) with code LCF/BQ/PI23/11970032. S.~M.~acknowledges support from the Flemish inter-university project IBOF/21/084 and would like to thank the L\'eon Rosenfeld Foundation for supporting his research visit to NBI. The work of L.~C.~was supported by the European Union’s H2020 ERC Consolidator Grant ``GRavity from Astrophysical to Microscopic Scales'' (Grant No. GRAMS-815673), the PRIN 2022 grant ``GUVIRP - Gravity tests in the UltraViolet and InfraRed with Pulsar timing'', and the EU Horizon 2020 Research and Innovation Programme under the Marie Sklodowska-Curie Grant Agreement No. 101007855. S.~H.~V.~acknowledges funding from the Deutsche Forschungsgemeinschaft (DFG): Project No. 386119226.
\end{acknowledgments}

\appendix

\section{Metric reconstruction}\label{app:perturbations}
We follow the analysis of \cite{Cano:2023tmv}, which at the same time made use of results from \cite{Dolan:2021ijg}. 
The perturbations to the rotating black hole solutions are described in terms of the four perturbed Teukolsky invariants $\{\delta\Psi_{0}\, , \delta\Psi_{4},  \delta\Psi_{0}^{*}\, , \delta\Psi_{4}^{*}\}$ (where $*$ denotes the Newman-Penrose conjugate) or equivalently, in terms of four Hertz potentials $\{\psi_{0}\, , \psi_{4},  \psi_{0}^{*}\, , \psi_{4}^{*}\}$.  The interplay between these variables is needed in order to evaluate the universal Teukolsky equation and reduced it to a decoupled radial equation. 
In the case of the Kerr solution in Einstein gravity, all these variables satisfy the Teukolsky equations of either spin $s=2$ (for the variables labeled with a 0) or spin $s=-2$ (for those labeled with a 4). Thus, we proceed to perform a mode expansion, and we can write the Hertz potentials as 
\begin{equation}
\begin{aligned}
\psi_0&=e^{-i\omega t+i m \phi} R_{2}(r)S_{2}(x)\, ,\\
\psi_0^{*}&=e^{-i\omega t+i m \phi} R_{2}^{*}(r)S_{-2}(x)\, ,\\
\psi_4&=e^{-i\omega t+i m \phi} \zeta^{-4} R_{-2}(r)S_{-2}(x)\, ,\\
\psi_4^{*}&=e^{-i\omega t+i m \phi} (\zeta^{*})^{-4} R_{-2}^{*}(r)S_{2}(x)\, ,
\end{aligned}
\end{equation}
and analogous expressions but with different radial functions hold for the Teukolsky variables.  Here $\zeta=r-i a x$ and $S_s(x)$ are the spin-weighted spheroidal harmonics, satisfying
\begin{align}\label{angularequation}\notag
&\frac{d}{dx}\left[(1-x^2)\frac{dS_{s}}{dx}\right]\\
&+\left[(a\omega)^2 x^2-2sa\omega x+ B_{lm}-\frac{(m+s x)^2}{1-x^2}\right]S_{s}=0\, ,
\end{align}
where $B_{lm}$ are the angular separation constants. We define $B_{lm}$ in a way that they are the same for $s=+2$ and for $s=-2$.
The variables $R_{s}$ and $R_{s}^{*}$ each satisfy the Teukolsky equation of spin $s$, 

\begin{equation}\label{radialeqs0}
\mathfrak{D}_{s}^2R_{s}=0\, ,\quad \mathfrak{D}_{s}^2R_{s}^{*}=0\, ,
\end{equation}
where $\mathfrak{D}_{s}^2$ was defined in \req{Dsopdef}.

On account of the equations \req{radialeqs0}, the radial variables are not independent. First, $R_{s}$ and $R_{s}^{*}$ satisfy the same equation, so they must be proportional to each other,
\begin{equation}\label{conjugaterelation}
R_{+2}^{*}(r)=q_{+2} R_{+2}(r)\, ,\quad R_{-2}^{*}(r)=q_{-2} R_{-2}(r)\, .
\end{equation}
To conclude this, one must bear in mind that they satisfy the same boundary conditions. 
The constants $q_{s}$ turn out to represent the polarization of the wave.  On the other hand, radial functions of different spin weight are related by the Starobinsky-Teukolsky (ST) identities, 
\begin{equation}\label{STidentities}
\begin{aligned}
R_{-2}&=C_{+2}\Delta^2\left(\mathcal{D}_{0}\right)^4\left(\Delta^2 R_{+2}\right)\, ,\\
R_{+2}&=C_{-2}\left(\mathcal{D}^{\dagger}_{0}\right)^{4}R_{-2}\, ,
\end{aligned}
\end{equation}
where $\mathcal{D}_{0}$ and $\mathcal{D}^{\dagger}_{0}$ are the operators 
\begin{equation}
\begin{aligned}
\mathcal{D}_{0}&=\partial_{r}+\frac{i\left(\omega(r^2+a^2)-m a\right)}{\Delta}\, ,\\
\mathcal{D}^{\dagger}_{0}&=\partial_{r}-\frac{i\left(\omega(r^2+a^2)-m a\right)}{\Delta}\, .
\end{aligned}
\end{equation}
The two proportionality constants $C_{\pm 2}$ satisfy
\begin{equation}\label{Cproduct}
C_{+2}C_{-2}=\frac{1}{\mathcal{K}^2}\, ,
\end{equation}
where 
\begin{equation}
\mathcal{K}^2=D_{2}^2+144 M^2 \omega^2\, ,
\end{equation}
and $D_{2}$ is the ST constant for the angular functions, given by\\

\begin{widetext}
\begin{equation}\label{D2value}
\begin{aligned}
D_{2}=\Big[&\left(8+6 B_{lm}+B_{lm}^2\right)^2-8 \left(-8+B_{lm}^2 (4+B_{lm})\right) m \gamma +4 \left(8-2 B_{lm}-B_{lm}^2+B_{lm}^3\dvvtag
+2 (-2+B_{lm}) (4+3 B_{lm})
m^2\right) \gamma ^2-8 m \left(8-12 B_{lm}+3 B_{lm}^2+4 (-2+B_{lm}) m^2\right) \gamma ^3\dvtag
+2 \left(42-22 B_{lm}+3 B_{lm}^2+8 (-11+3
B_{lm}) m^2+8 m^4\right) \gamma ^4\dvtag-8 m \left(3 B_{lm}+4 \left(-4+m^2\right)\right) \gamma ^5
+4 \left(-7+B_{lm}+6
m^2\right) \gamma ^6-8 m \gamma ^7+\gamma ^8\Big]^{1/2}\, .
\end{aligned}
\end{equation}
\end{widetext}
with $\gamma=a\omega$.
Finally, using all these relations one can show that the Teukolsky variables resulting from these Hertz potentials are given by
\begin{equation}\label{metricweylrelation}
\delta\Psi_{2-s}=P_{s}\psi_{2-s}\, ,\quad \delta\Psi_{2-s}^{*}=P_{s}^{*}\psi_{2-s}^{*}\, .
\end{equation}
where the proportionality constants $P_{s}$, $P_{s}^{*}$ read
\begin{align}\label{eq:Pconstants}
P_{s}&=\frac{1}{2}+\frac{i s}{48 M \omega }\left(D_2 q_{s}- 2^s  q_{-s} C_{s}\mathcal{K}^2\right)\, ,\\
P_{s}^{*}&=\frac{1}{2}+\frac{i s}{48 q_{s}M \omega }\left(D_2-2^s C_{s} \mathcal{K}^2\right)\, ,
\end{align}
with $s=\pm 2$. 
All of the relationships \req{conjugaterelation}, \req{STidentities} and \req{metricweylrelation} allow us to express any of these variables in terms of any of the other.  In fact, since the metric perturbation is determined by the Hertz potentials, we can write any perturbed quantity --- frame, spin connection, Riemann curvature --- in terms of any of the radial functions and the constants $q_s$ and $C_s$.

Now, these relationships no longer hold in the case of higher-derivative gravity, since all of them will undergo perturbative corrections.  However, it suffices to know them in the case of GR, since we only need to use them in the terms of the universal Teukolsky equations that are proportional to the higher-derivative corrections, so that the corrections to  \req{conjugaterelation}, \req{STidentities} and \req{metricweylrelation}  would be a second-order effect.  Using these results, one can proceed to evaluate the universal Teukolsky equations linearized over the background of a rotating black hole, and express the result in terms of the radial functions $R_{s}$ and $R_{s}^{*}$. 
The resulting equation however is non-separable, and thus one has to do a mode expansion including all the different $l$ modes. The equation is then projected onto the spheroidal harmonics, giving an infinite system radial equations that couple different $l$ modes. However, the equation for the dominant mode in the expansion is decoupled and determines the QNM frequency. We explained this in the text around Eq.~\req{deltapsiexpansion}.

During the process, terms with more than two derivatives of the radial functions can be reduced by using the zeroth-order Teukolsky equation.

\bibliographystyle{apsrev4-2} % Tell bibtex which bibliography style to use
\vspace{1cm}
\bibliography{Gravities} % Tell bibtex which .bib file to use (this one is some example file in TexLive's file tree)

\end{document}